\documentclass[proof]{WileyASNA-v1}

\usepackage{graphicx}	
\usepackage{amsmath}	
\usepackage{hyperref}
\usepackage{tabularx}
\usepackage{subcaption}
\usepackage{ulem}

\defcitealias{2022MNRAS.517..484N}{Paper\,I}
\defcitealias{2021MNRAS.505.5978V}{VB21}

\graphicspath{{images/}}
\articletype{Original Article}%

\received{4 January 2023}
\revised{15 February 2023}
\accepted{21 February 2023}

\begin{document}

\title{Photometry and astrometry with \textit{JWST} -- II.\\ NIRCam distortion correction}

\author[1,2]{M. Griggio*}
\author[1,3]{D. Nardiello*}
\author[1]{L. R. Bedin*}

\authormark{M. Griggio, D. Nardiello \& L. R. Bedin}

\address[1]{\orgdiv{Istituto Nazionale di Astrofisica}, \orgname{Osservatorio Astronomico di Padova}, \orgaddress{\state{Vicolo dell'Osservatorio 5, Padova, IT-35122}, \country{Italy}}}

\address[2]{\orgdiv{Universit\`a di Ferrara}, \orgname{Dipartimento di Fisica}, \orgaddress{\state{Via Giuseppe Saragat 1, I-44122, Ferrara}, \country{Italy}}}

\address[3]{\orgdiv{CNRS, CNES}, \orgname{Aix Marseille Univ., LAM}, \orgaddress{\state{Marseille}, \country{France}}}

\corres{*E-mails: (massimo.griggio, luigi.bedin, domenico.nardiello)@inaf.it}

\abstract{
In preparation to make the most of our own planned \textit{James Webb Space Telescope} investigations, 
we take advantage of publicly available calibration and early-science observations to independently 
derive and test a geometric-distortion solution for NIRCam detectors. 
Our solution is able to correct the distortion to better than $\sim$0.2\,mas.
Current data indicate that the solution is stable and constant over the investigated filters, temporal coverage, 
and even over the available filter combinations. 
We successfully tested our geometric-distortion solution in three cases: (i) field-object decontamination of M\,92 field; (ii) estimate of internal proper motions of M\,92; and (iii) measurement of the internal proper motions of the Large Magellanic Cloud system.
To our knowledge, the here-derived geometric-distortion solution for NIRCam is the best 
available and we publicly release it, as many other investigations 
could potentially benefit from it.
Along with our geometric-distortion solution, we also release a \texttt{Python} tool to convert the raw-pixels coordinates of each detector into distortion-free positions, and also to put all the ten detectors of NIRCam into a common reference system. 
}


\maketitle


\section{Introduction, Observations, Data-Reduction}
\label{sec:obs}

\begin{figure*}
    \centering
    \includegraphics[width=\columnwidth]{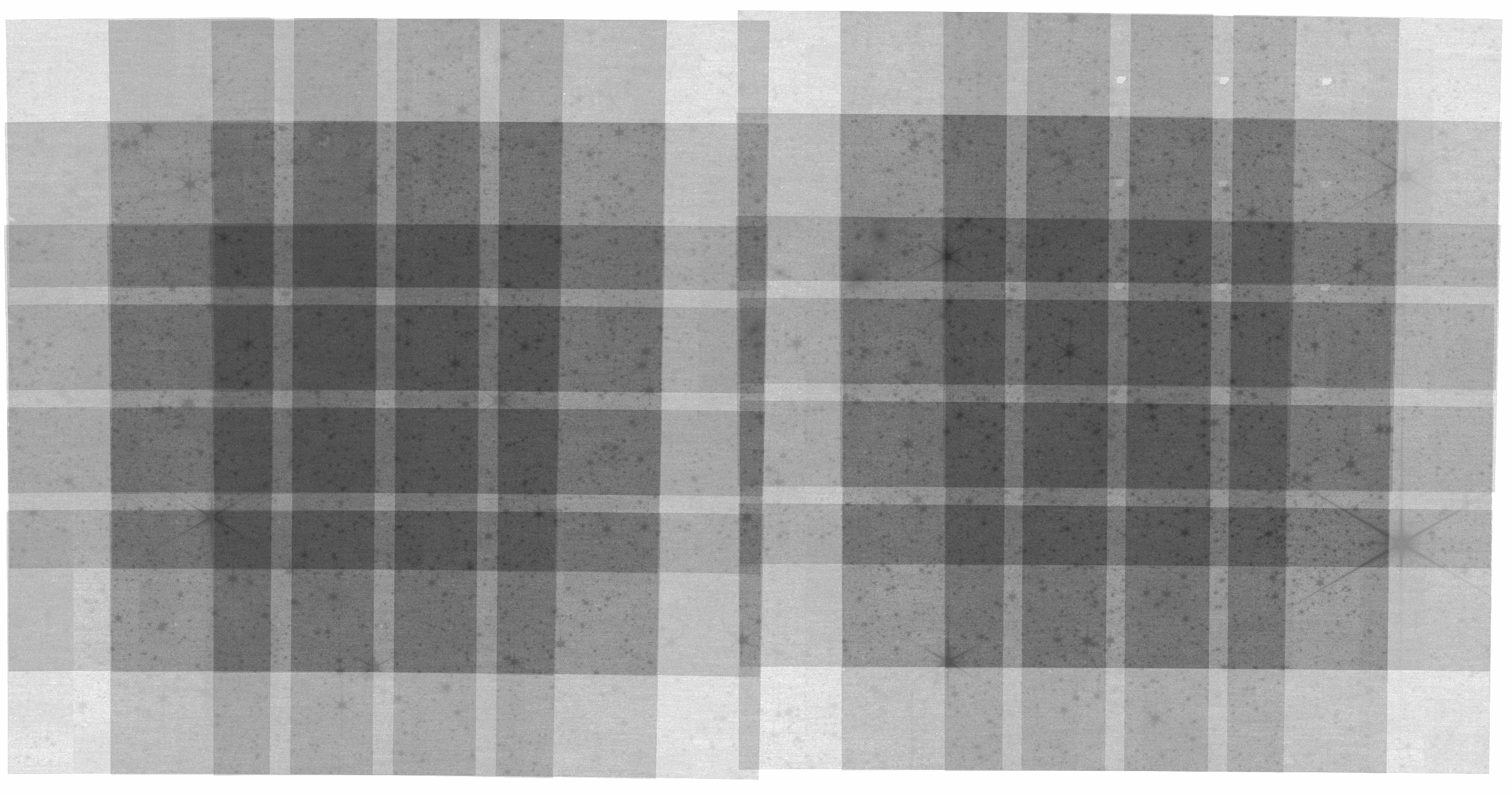} 
    \includegraphics[width=\columnwidth]{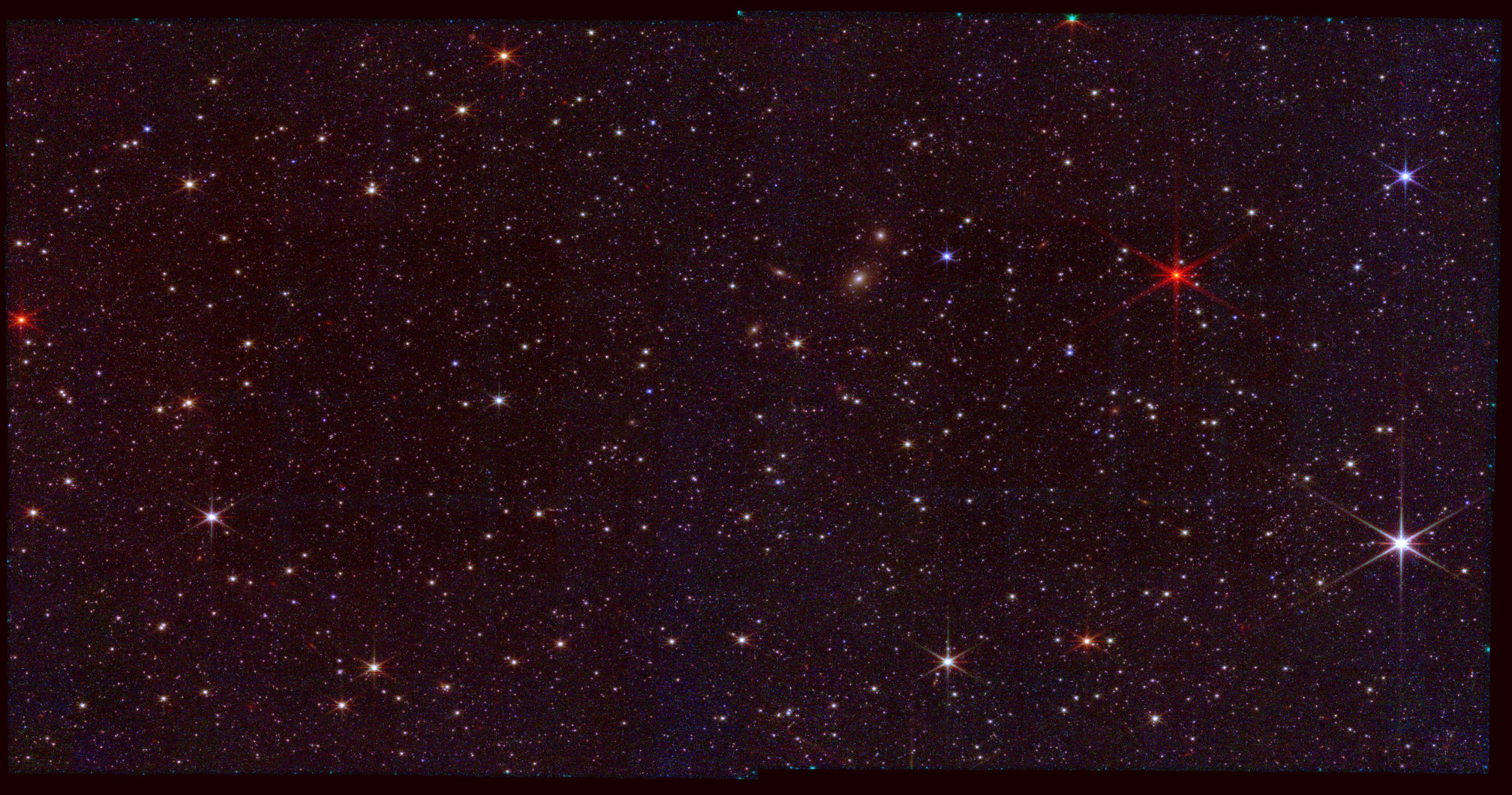} \\
    \caption{
\textit{(Left:)} depth-of-coverage of the 9 pointings for each
considered filter in the SW channel. 
The studied region in the LMC cover about $6^\prime\times3^\prime$, and shows  
large overlaps between the SW's detectors. 
\textit{(Right:)} a three-colour view of the region, where F090W, F150W and F444W were used for blue, green and red colour, respectively.}
    \label{fig:dither}
\end{figure*}

\begin{figure*}
    \centering
    \includegraphics[width=.95\textwidth]{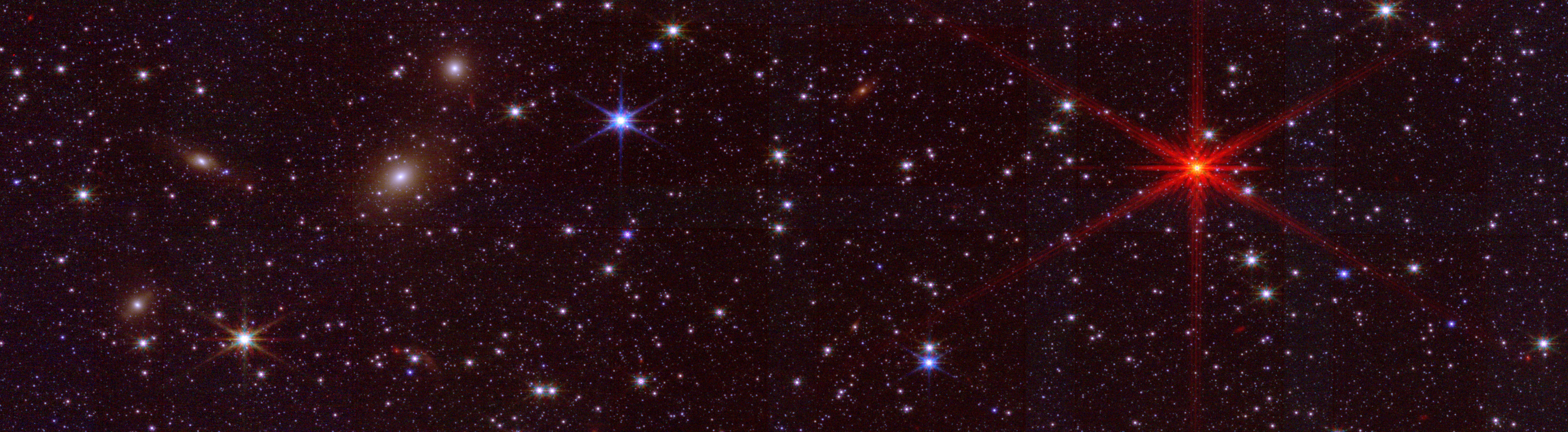} 
    \caption{
To give an idea of the richness of isolated well-measurable sources in the field  
we show a zoom-in of a portion of $\sim150^{\prime\prime}\times44^{\prime\prime}$, around 
the brightest and reddest source ({\it Gaia}~DR3~\texttt{4657988450340570624}, \texttt{2MASS~05212923-6927554}, \texttt{WISE~J052129.23-692755.4}) visible in right panel of Figure\,\ref{fig:dither}(a red super-giant 
belonging to LMC classified as an extreme AGB star by \citealt{2011AJ....142..103B}) . 
    } 
    \label{fig:zoom}
\end{figure*}

The characterisation of the geometric distortion (GD) of an imager 
is of paramount importance to assess its use for high-precision astrometry.  
This is particularly important in the case of cameras of an out-of-atmosphere, 
brand-new instrument, such as the \textit{James Webb Space Telescope (JWST)}, 
arguably the world-wide most-important astronomical facility.

In this work, we made use of part of \textit{JWST} public data collected with the 
\textit{Near Infrared Camera}\footnote{\url{https://jwst-docs.stsci.edu/jwst-near-infrared-camera}} 
(NIRCam) under the Cycle\,1 Calibration Program 
\href{https://www.stsci.edu/jwst/science-execution/program-information.html?id=1476}{1476}
(PI: M.~Boyer) to derive its GD correction. 
While standard pipeline products for GD corrections of \textit{JWST}’s cameras are expected to be released in the future by other teams, we provide with this work the first, independent, 
documented and 
publicly available GD correction that allows an accuracy of $\sim 0.2$ mas on the stellar positions.

This paper is part of a series to build-up our capabilities to 
obtain \textit{state-of-the-art} imaging astrometry and photometry with \textit{JWST}.
This is a necessary task for us to properly prepare and 
maximise the scientific returns of our planned (March 2023) 
proprietary \textit{JWST} observations
(\href{https://www.stsci.edu/jwst/science-execution/program-information.html?id=1979}{GO-1979}, PI: Bedin). 

In our first paper 
\cite{2022MNRAS.517..484N} (hereafter Paper\,I), we 
describe the procedure to derive high-accuracy point-spread functions (PSFs) 
for NIRCam in some filters, 
an essential step to derive high-precision photometry, especially in crowded environments.
We made those 
PSFs publicly available\footnote{\url{https://web.oapd.inaf.it/bedin/files/PAPERs_eMATERIALs/JWST/Paper_01/}}. In this second paper, we also release to the public 
our GD correction of NIRCam. 

Calibration Program 1476 will derive the GD for both channels of NIRCam
by observing with \textit{JWST} the Large Magellanic Cloud (LMC) calibration field observed 
multiple times with the \textit{Hubble Space Telescope (HST)}. 
This field is centred at $\alpha=80^\circ\!\!.49030$, $\delta=-69^\circ\!\!.49816$, 
and is described in the {\it JWST} technical report \cite{2021jwst..rept....12A}.

However, in the present work we will not make use of this \textit{HST} astrometric 
catalogue to derive our GD correction of NIRCam.  
Instead, we will make use just of the new \textit{JWST} observations to self-calibrate (i.e. to calibrate without exploiting observations of standard
astrometric  fields), 
leveraging the existing {\it Gaia} Data\,Release\,3 (DR3) \citep{2022arXiv220800211G}
to constrain the linear terms of our GD solution.
In this sense, our work is an independent analysis and solution of the NIRCam GD, to be compared in the future with those that will be released by the instrument team.

We employed the set of images collected with the Short Wavelength (SW) channel in F090W, F150W, and F150W2 filters, and with the Long Wavelength (LW) channel in F277W and F444W. 
We also, test the derived geometric distortion solution in the available filter combos: F150W2+F162N, F150W2+F164N,
F444W+F405N, F444W+F466N and F444W+F470N.

\begin{table}
    \caption{Log of the observations employed in this work.}
    \label{tab:obs}
    \begin{tabularx}{\columnwidth}{lllll}
        \hline \hline
        Filter & Pupil & $t_{\rm exp}$\,[s]  & Readout pattern & $N_{\rm exp}$\\
        \hline
        F090W  & CLEAR & 21.474         & \texttt{RAPID}    & 9  \\
        F150W  & CLEAR & 21.474         & \texttt{RAPID}    & 9  \\
        F150W2 & CLEAR & 21.474         & \texttt{RAPID}    & 9  \\
        F150W2 & F162M & 85.894         & \texttt{BRIGHT2}  & 9  \\
        F150W2 & F164N & 118.104        & \texttt{BRIGHT1}  & 18 \\
        F277W  & CLEAR & 21.474         & \texttt{RAPID}    & 9  \\
        F444W  & CLEAR & 21.474         & \texttt{RAPID}    & 9  \\
        F444W  & F405N & 118.104        & \texttt{BRIGHT1}  & 9  \\
        F444W  & F466N & 257.682        & \texttt{SHALLOW4} & 9  \\
        F444W  & F470N & 257.682        & \texttt{SHALLOW4} & 9  \\
        \hline
    \end{tabularx}
\end{table}

For each filter, {\it JWST} observed the field with 9 different pointings 
in such a way a given star is placed in 9 different positions of a detector (Fig.\,\ref{fig:dither}). Each pointing is an exposure obtained with a single integration. In Table~\ref{tab:obs} we report all the observations used in this work.

\begin{figure}
    \centering
        \includegraphics[width=\columnwidth]{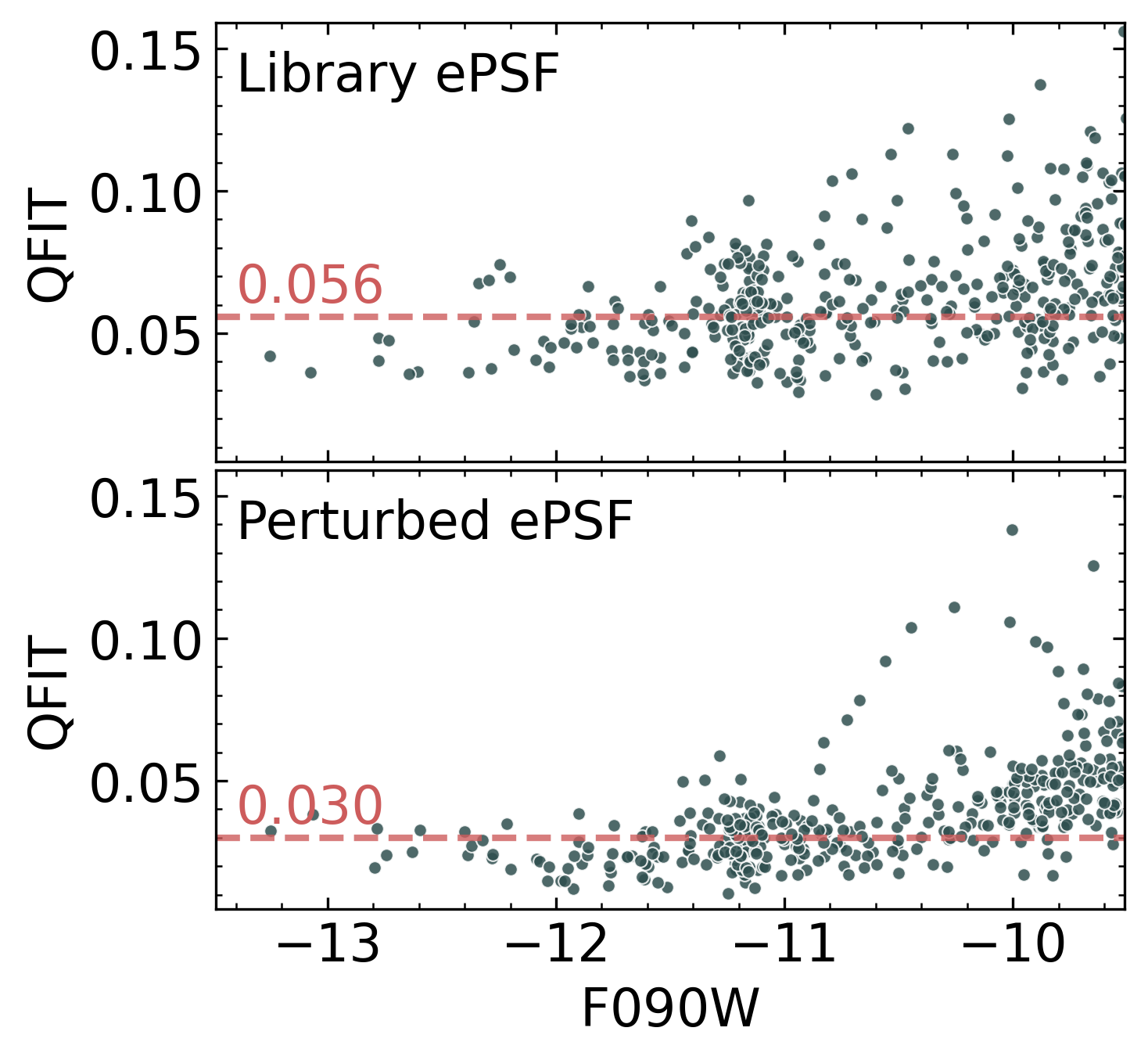}
    \caption{
    The quality-of-fit (\texttt{QFIT}) distribution before (top panel) and after (bottom panel) the ePSF perturbation. The median \texttt{QFIT} value decreases from 0.056 to 0.030, with an improvement of $\sim$50\,\% in the PSF fitting. The figure refers to an image in F090W filter, namely \texttt{jw01476001003\_02101\_00001}, module B, detector 2.}
    \label{fig:qfit}
\end{figure} 

We extracted catalogues of positions and fluxes for point sources from the NIRCam calibrated images\footnote{\url{https://jwst-pipeline.readthedocs.io/}} (\texttt{\_cal}) by adopting the procedure described in 
\citetalias{2022MNRAS.517..484N}. 
Briefly, for each image, we used a list of bright, isolated, unsaturated stars to perturb the $5 \times 5$ library effective PSFs (ePSFs) obtained in \citetalias{2022MNRAS.517..484N}, in such a way as to take into account the time variations of the {\it JWST} ePSFs. Briefly, the software we adopted measured the flux and the positions of the stars by fitting the library ePSFs, subtracted the models of these stars from the image, and calculated the average of the normalized residuals that are finally added to the library ePSFs. The routine carried out 11 rounds if iterations, and at each iteration the residuals are used to adjust the last obtained PSFs.
Figure\,\ref{fig:zoom} shows a zoom-in of the studied region at a meaningful scale to display individual sources;
it is representative of the entire field, which is rather homogeneous and rich in bright sources, relatively isolated, and well-measurable.

In Fig.~\ref{fig:qfit} we show why it is important to perturb the library ePSF. 
In the top panel, the quality-of-fit parameter (\texttt{QFIT}) obtained employing the library ePSFs (from \citetalias{2022MNRAS.517..484N}) 
is shown as a function of the instrumental magnitudes 
($m_{\rm instr} = -2.5\log{\Sigma ({\rm counts})}_{\rm used\,pixels}$). 
The parameter \texttt{QFIT} essentially quantifies the difference between the adopted ePSF model and the observed stars on the images. In the bottom panel, the same PSF diagnostic is shown when the perturbed ePSFs are used. 
In this case, the \texttt{QFIT} parameter significantly decreases, getting closer to zero, so the PSF better resembles the real stars. 
This translates into improved astrometry, photometry, and source separations. 

We adopted the software 
\texttt{img2xym} 
developed by J.\,Anderson (\citealt{2006A&A...454.1029A}), and adapted to the NIRCam images, to extract positions and fluxes of the stars by fitting the perturbed ePSFs; we searched for sources  with a total flux $\ge 50$ counts and whose peaks are isolated at least 5 pixel from the closest brighter pixel. The software identifies the peaks that satisfy these criteria among the image, and for each of them, it fits the local ePSF obtained by the bi-linear interpolation of the 4 closest perturbed ePSF of the grid. Through a chi-square minimisation, it measures the position and flux of each source. We refer the reader to \citetalias{2022MNRAS.517..484N} for a more detailed description of the data reduction.

\section{Geometric distortion correction}
\begin{figure*}
    \centering
    \begin{subfigure}{.49\textwidth}
        \centering
        \includegraphics[width=\textwidth]{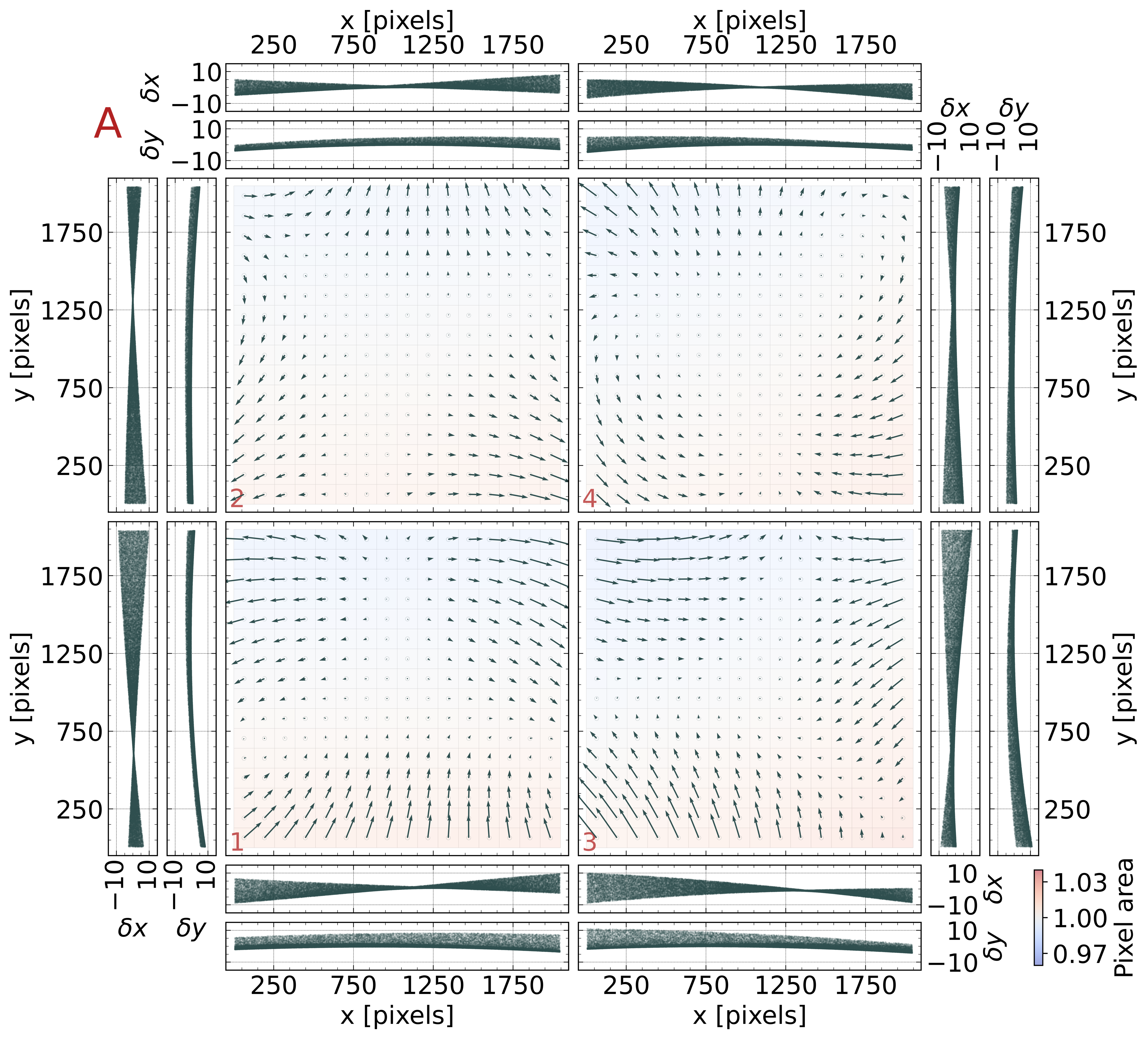}
        \caption{}
        \label{fig:gd_modA_SW}
    \end{subfigure}
    \hfill
    \begin{subfigure}{.49\textwidth}
        \centering
        \includegraphics[width=\textwidth]{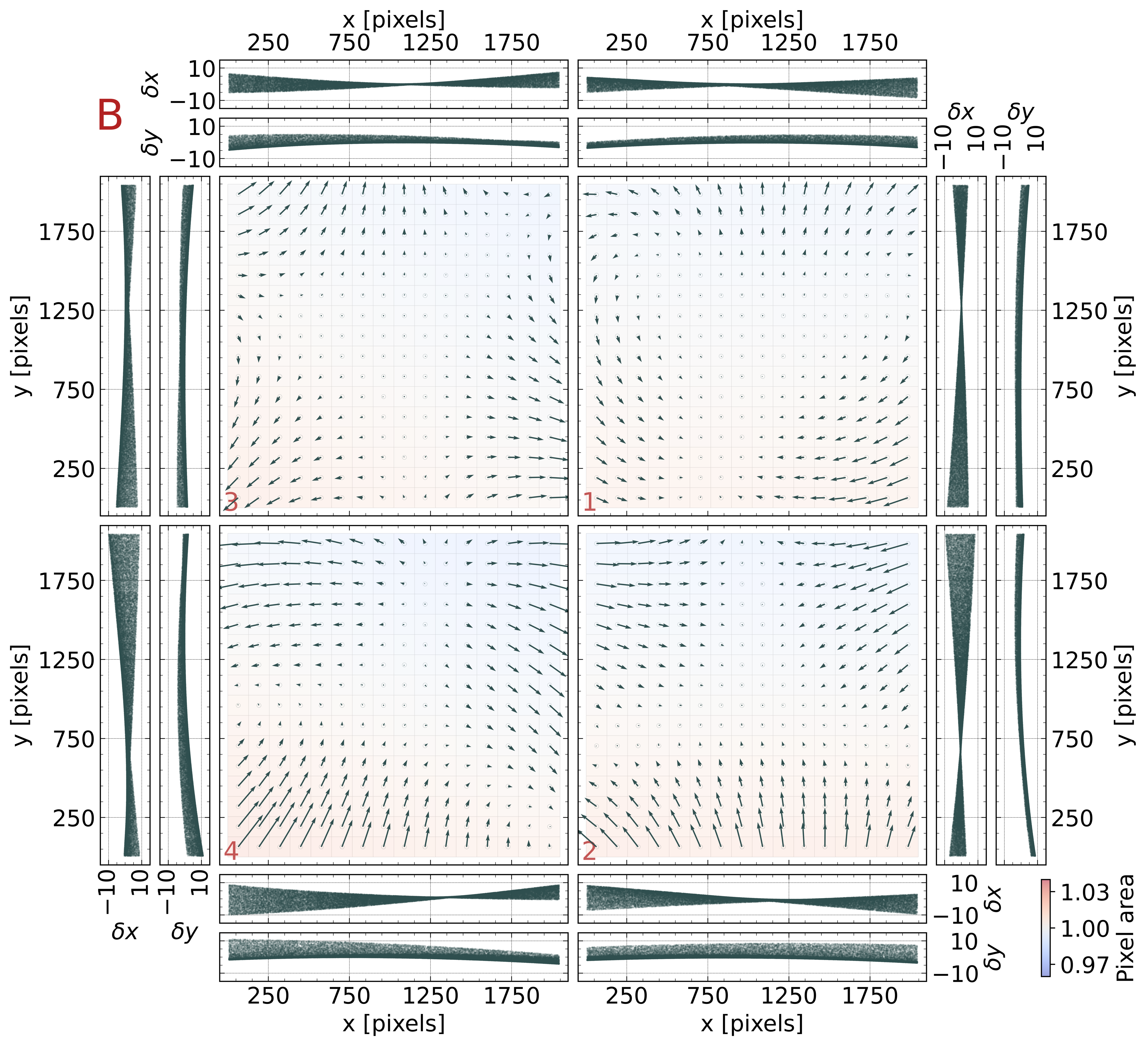}
        \caption{}
        \label{fig:gd_modB_SW}
    \end{subfigure}
    \hfill
    \centering
    \begin{subfigure}{.49\textwidth}
        \centering
        \includegraphics[width=\textwidth]{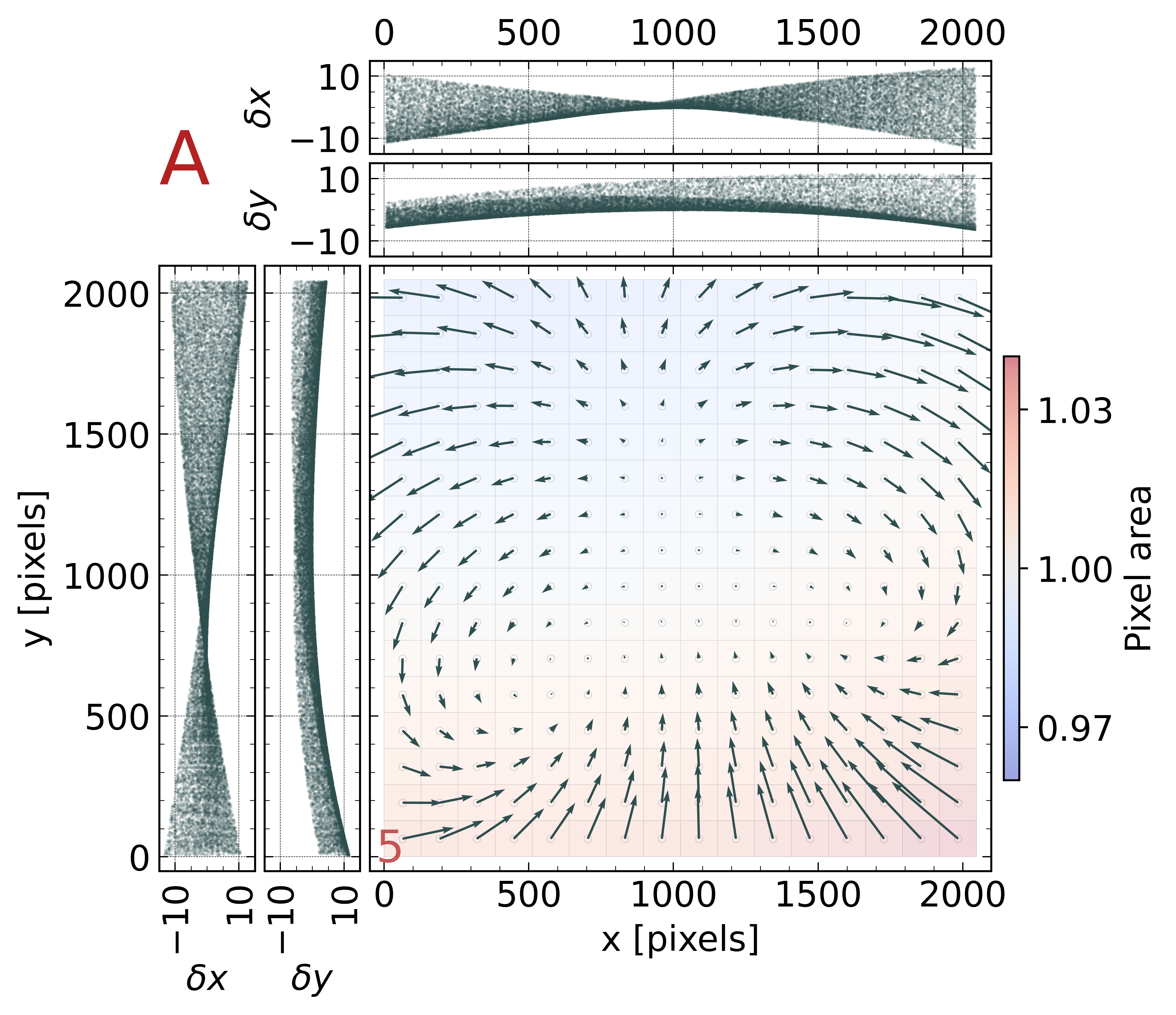}
        \caption{}
        \label{fig:gd_modA_LW}
    \end{subfigure}
    \hfill
    \begin{subfigure}{.49\textwidth}
        \centering
        \includegraphics[width=\textwidth]{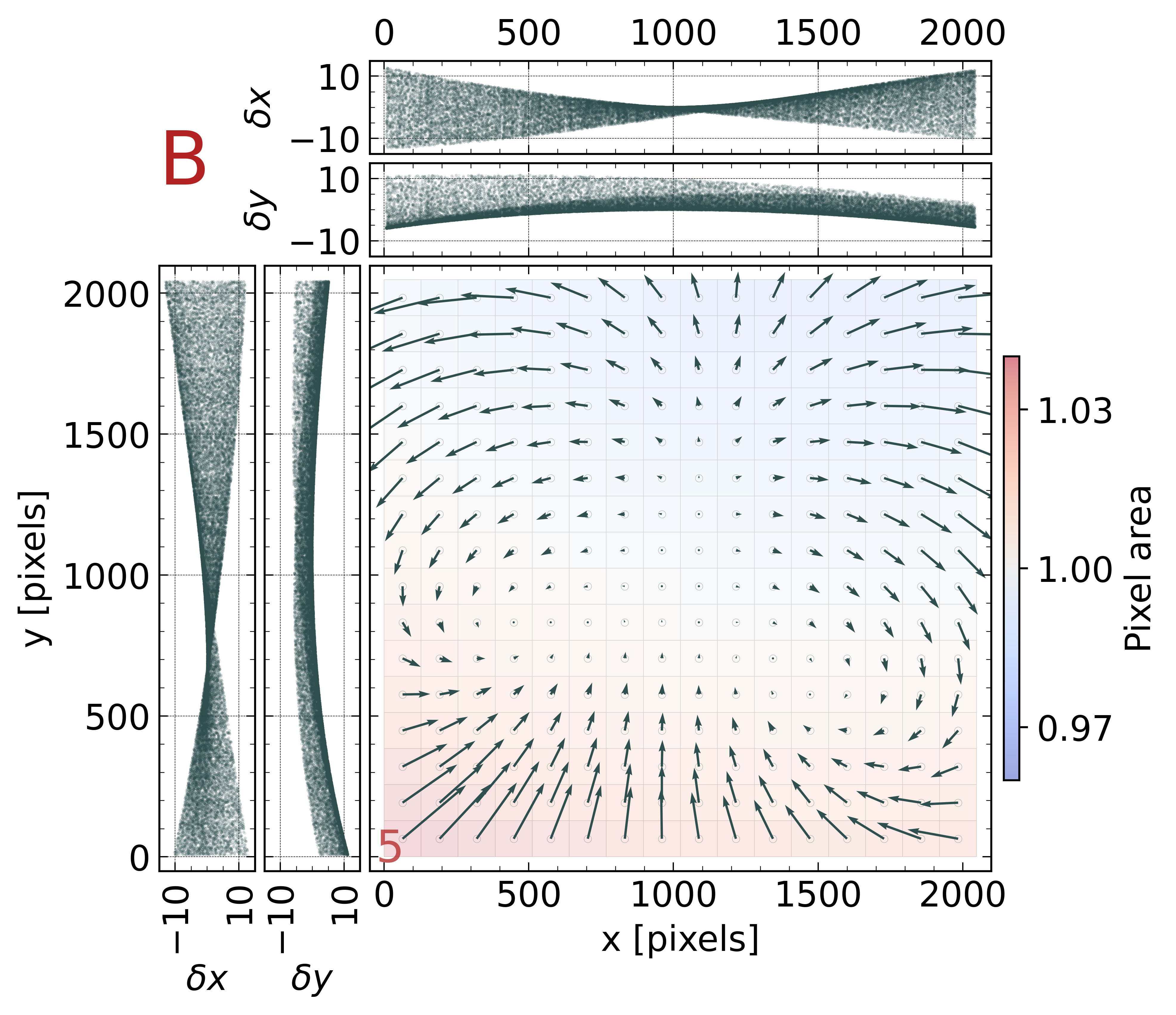}
        \caption{}
        \label{fig:gd_modB_LW}
    \end{subfigure}
    \caption{
    Geometric distortion map of NIRCam short (top) and long (bottom) wavelength channel for both  
    module A (left) and module B (right); 
    detector number is shown
    in red on the bottom left corner of each vector plot. 
    The size of the residual vectors is magnified by a factor 20. For each detector we also show 
    the single residual trends along $x$ and $y$ axes. Units are in raw NIRCam pixels.
    The colour map represents the pixels' area variation across the detectors (see text).} 
    \label{fig:gd_jwst}
\end{figure*}

Our derivation of the GD correction for NIRCam
followed the empirical approach and procedure 
successfully applied to derive the GD correction 
of many other cameras at the focus 
of space- and ground-based telescopes 
\citep{
2003PASP..115..113A,
2006A&A...454.1029A,
2009PASP..121.1419B,
2011PASP..123..622B,
2014A&A...563A..80L,
2015MNRAS.450.1664L,
2019MNRAS.484.5530K}.

Our GD solution is derived independently for each of the ten 2048$\times$2048\,pixels 
NIRCam detectors (8 for SW, and 2 LW), and it is made up of three parts. 
First, a backbone third-order 
polynomial (Section \ref{sec:poly}) derived through a self-calibration procedure; 
second, a first-order polynomial derived by exploiting the {\it Gaia}\,DR3 reference system, 
to fix the linear terms of the GD (Section \ref{sec:poly1}), and 
third, a fine-scale table that accounts for spatial high-frequency systematic residuals 
that the polynomial correction can not absorb (Section \ref{sec:tab}).
More details on self-calibration of the GD can be found
in the work by \cite{2003PASP..115..113A}, of which we will give a brief description in next sub-section.

The procedure that we will describe in the next sections have been
applied independently to each detector of both modules (A and B).
The final distortion map is shown in Figure \ref{fig:gd_jwst}.
The colour map represents the pixels' area
variation across the detectors due to the GD, which 
is relevant to show for those investigations dealing with surface brightness. 
Each $128\times128$\,pixels region in the vector plots of Figure \ref{fig:gd_jwst} is coloured according to 
the ratio between the GD corrected area and the raw area.
We computed the GD corrected area using the corrected positions of the corners of
each region, thus the value represents the mean area
variation of the pixels in that region.

\subsection{Polynomial correction}
\label{sec:poly}

The polynomial GD solution is represented by a
third-order polynomial; we checked that higher-orders
(fifth and seventh) do not provide better results.
We chose the pixel $(x_0,y_0)=(1024, 1024)$  in each
detector as a reference position and solved for the distortion
with respect to it, using the normalised coordinates
$\left(\tilde{x},\tilde{y}\right)=\left(\frac{x-x_0}{x_0}, \frac{y-y_0}{y_0}\right)$ \citep[cfr.][]{2003PASP..115..113A,
2011PASP..123..622B}.

To derive the polynomial coefficients we performed
a series of iterations in which we alternate two main tasks: 
building the master frame, which will be the temporary reference system closer to
the distortion-free solution than the previous iteration,
and calculating the residuals between the positions of
the sources as measured on the master frame and those
measured in the raw catalogues. These residuals are then
used to derive the polynomial coefficients.  
The polynomial correction is performed as follows:
\begin{itemize}
\item[--] We selected the sources in each catalogue with instrumental
magnitude in the range $-12.5<m_{\rm instr}<-8$ and with \texttt{QFIT}
lower than 0.1, to avoid artefacts, saturated and poorly measured stars which would
affect the distortion solution.
\item[--] We conformally transformed the positions of each star in each catalogue
into the reference system of the central dither, and we built a master frame by averaging the positions and fluxes of the sources that are measured
in at least three exposures.
\item[--] At this point, we computed the conformal transformations ($T$) between stars
in each catalogue and the master frame.
\item[--] The inverse transformation ($T^{-1}$) is then used to compute the positions of the 
master frame's stars in the raw-coordinate system of each image, that are 
then cross-identified with the closest source
after applying the inverse GD correction
derived in the previous iteration (which, of course, at the first iteration is
equal to the identity).
Each such cross-identification generates a pair of positional residuals 
$\delta x = x_{\mathrm{raw}} - X^{T^{-1}\,\circ\,GD^{-1}}$ and 
$\delta y = y_{\mathrm{raw}} - Y^{T^{-1}\,\circ\,GD^{-1}}$, where
$(x_{\mathrm{raw}},y_{\mathrm{raw}})$ and $(X,Y)$ are the coordinates in the 
raw catalogues and the master frame reference systems respectively.
\item[--] We performed a least-square fit of the residuals
to obtain the coefficients 
for the two third-order polynomials that are added to those derived at the
previous iteration. To ensure convergence, the 75 percent 
of the correction is then applied to all stars’ positions.  
\end{itemize}
The procedure is iterated over up to 45 iterations, until convergence is reached, 
starting from the corrected catalogues, each time refining the master frame and the polynomial coefficients.
At the end of the procedure, we had a set of coefficients for each filter. 
The polynomials derived independently for each filter turned out to be in agreement 
within the uncertainties, 
therefore we computed a weighted mean (using the inverse of the errors
on the coefficients as weights) 
to get a single final polynomial for each individual detector.
While the polynomials for filter combinations
were marginally in agreement with those obtained for single filters, 
they were not used to compute the average polynomials. 
%
\subsection{GD linear terms}
\label{sec:poly1}
%
\begin{figure}
    \centering
    \includegraphics[width=\columnwidth]{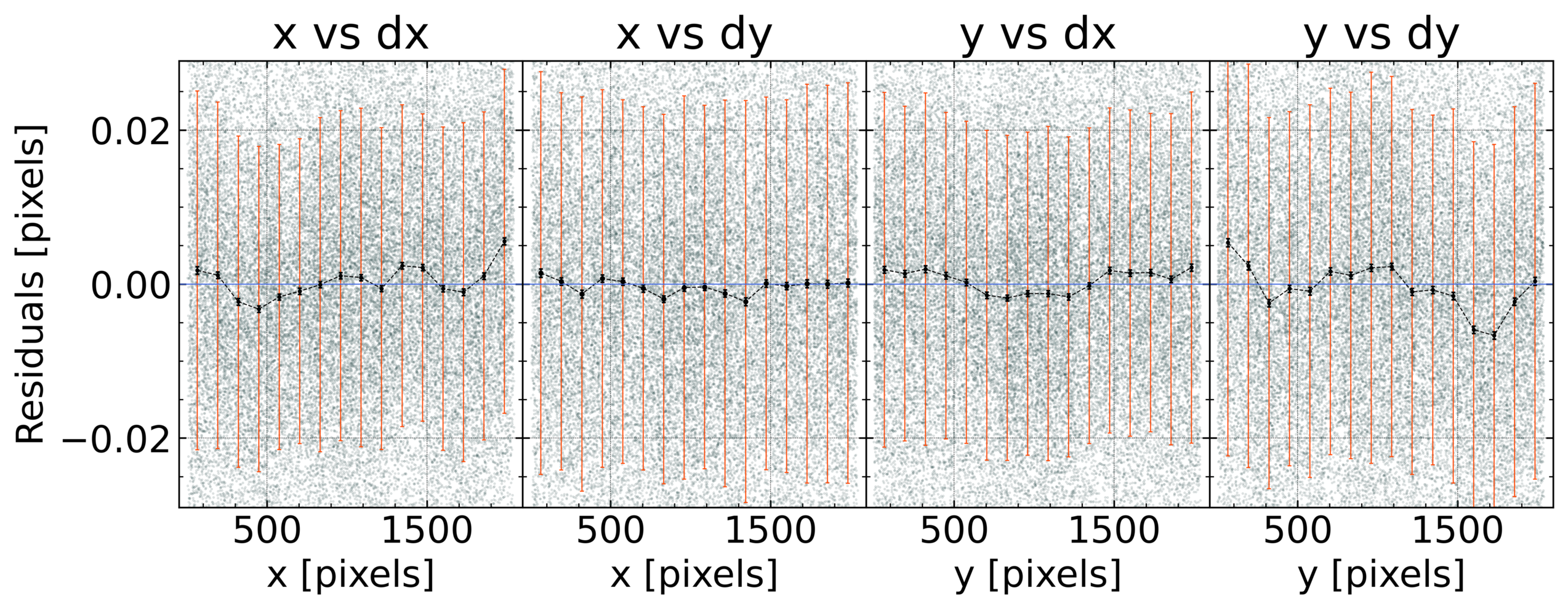}
    \caption{Positional residuals given by
    the inter-comparison between all the dithered exposures (for detector A1)
    in F150W and F150W2 filters after the two polynomial have been applied.
    The black dots represent the mean residual in each slice of 128 pixels.
    The red error bars are calculated as $\sigma=68.27^{\rm th}$ percentile of
    the residual distribution (after a $3\sigma$-clipping),
    and the black error bars are $\sigma/\sqrt{n-1}$, with $n$ the number of
    points used to compute the mean.}
    \label{fig:gd_resA1_p1}
\end{figure}

So far, the first epoch of calibration program 1476 has observations collected
at one single orientation of the telescope. 
This makes it not possible to solve for the
linear terms of the GD \citep{2003PASP..115..113A}.
For this reason, we will make leverage of the existing astrometric flat field 
provided by {\it Gaia} DR3 to perform this task. 

While common sources are very few, faint and poorly measured,
we need only 3 stars, in principle, to fix these linear terms, 
as the most general linear transformation has only 6 parameters 
(therefore the 2D positions of three stars would be sufficient).

Nevertheless, in each detector, there are always at least 350 stars
in common between {\it Gaia} and NIRCam observations of program 1476
for the SW channel, and at least 1200 for LW channel, more than enough
for our purposes.

We then proceeded in the same way as described in the previous section,
but this time using {\it Gaia} (projected onto the tangent plane of each
exposure) as a master frame, starting with the catalogues
corrected with the third-order polynomial, and using all the filters together.
We needed 10 iterations to reach convergence.

The residuals between the inter-comparison between all the dithered exposures
for detector A1 are shown 
in Figure \ref{fig:gd_resA1_p1}:
we notice the presence of small spatial residuals, that we wanted to remove.
We corrected these systematics with a lookup table as described in
the next section.

\subsection{Fine-structure table of residuals}
\label{sec:tab}
\begin{figure}
    \centering
    \includegraphics[width=\columnwidth]{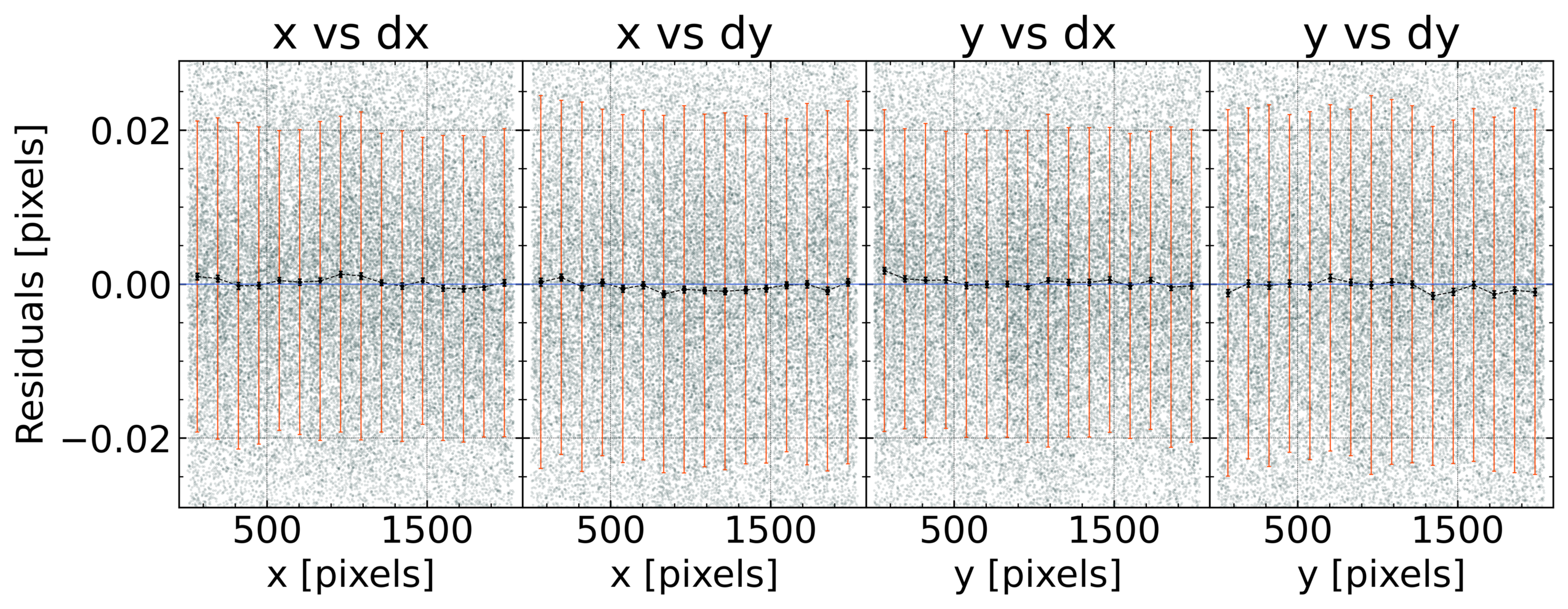}
    \caption{As Fig. \ref{fig:gd_resA1_p1} but after applying also the fine-scale table correction. 
    These black dots are our internal errors, which 
    are always smaller than
    20\,$\mu$as, and are the 
    formal uncertainty of our GD correction; whereas the larger error bars (in 
    red) show the positional random error for the individual ``typical'' source. 
    }
    \label{fig:gd_resA1_tab}
\end{figure}

The systematic residuals observed in
Fig.\,\ref{fig:gd_resA1_p1}
could not be removed with the polynomial corrections. 
Additional iterations did not provide any improvements. 
For this reason, we decided to proceed with a fine-structure table
of residuals. 

We followed two different procedures for the SW channel and the LW channel; 
for the SW channel we employed again a self-calibration procedure.
We started from the catalogues corrected with the two polynomials 
and followed the first four
steps of the bullet list in Section \ref{sec:poly}.
We then divided the residuals into a lookup table of $16\times16$ cells in $x$ and $y$.
To each cell, we assigned a residual in $x$ and $y$ using the $3\sigma$-clipped mean
of the residuals in that cell.
The positions of the stars are then corrected with the residual calculated with
a bilinear interpolation
of the four most adjacent cells \citep[cfr.][]{2014A&A...563A..80L}.
This procedure is iterated over 10 times to converge. 
The systematic residuals were successfully corrected;
after the correction, the inter-comparison of corrected frames
is consistent to the sub-mas level (Figure \ref{fig:gd_resA1_tab}, 
assuming a pixel scale of 
31.23\,mas, see Section \ref{sec:meta}).

We applied the same self-calibration procedure to the LW channel, unsuccessfully.
After 10 iterations, the residuals between the inter-comparison of dithered exposures
did not show any clear trend. However, the comparison of these positions 
with the {\it Gaia} catalogue showed a global trend in the residual distribution: 
we suspect that the data are insufficient for a self-calibration of the GD for the LW channel.
Indeed, the dither pattern (which is the same for both channels) offers larger inter-comparison overlaps for SW than for LW.

Therefore, we exploited our just corrected SW catalogues to build a distortion-free 
master frame,
on which we can calibrate also the LW channel. We proceeded with the same steps that we
followed to derive the SW channel lookup table, but this time as a master frame we
employed the one built employing the SW corrected catalogues.
Adopting this procedure, also in this case, 10 iterations were sufficient to reach 
convergence.

This is the final step that concludes the derivation of our GD correction for the NIRCam detectors. 
In Sect. \ref{sec:sum} we will give details about the \texttt{Python} routine, which we release as 
electronic material part of this publication, that will enable readers to transform the 
raw pixel coordinate of each of the 10 individual detectors of NIRCam into a distortion-free frame. 

\subsection{\textit{Gaia} validation}
%
\begin{figure}
    \centering
    \includegraphics[width=\columnwidth]{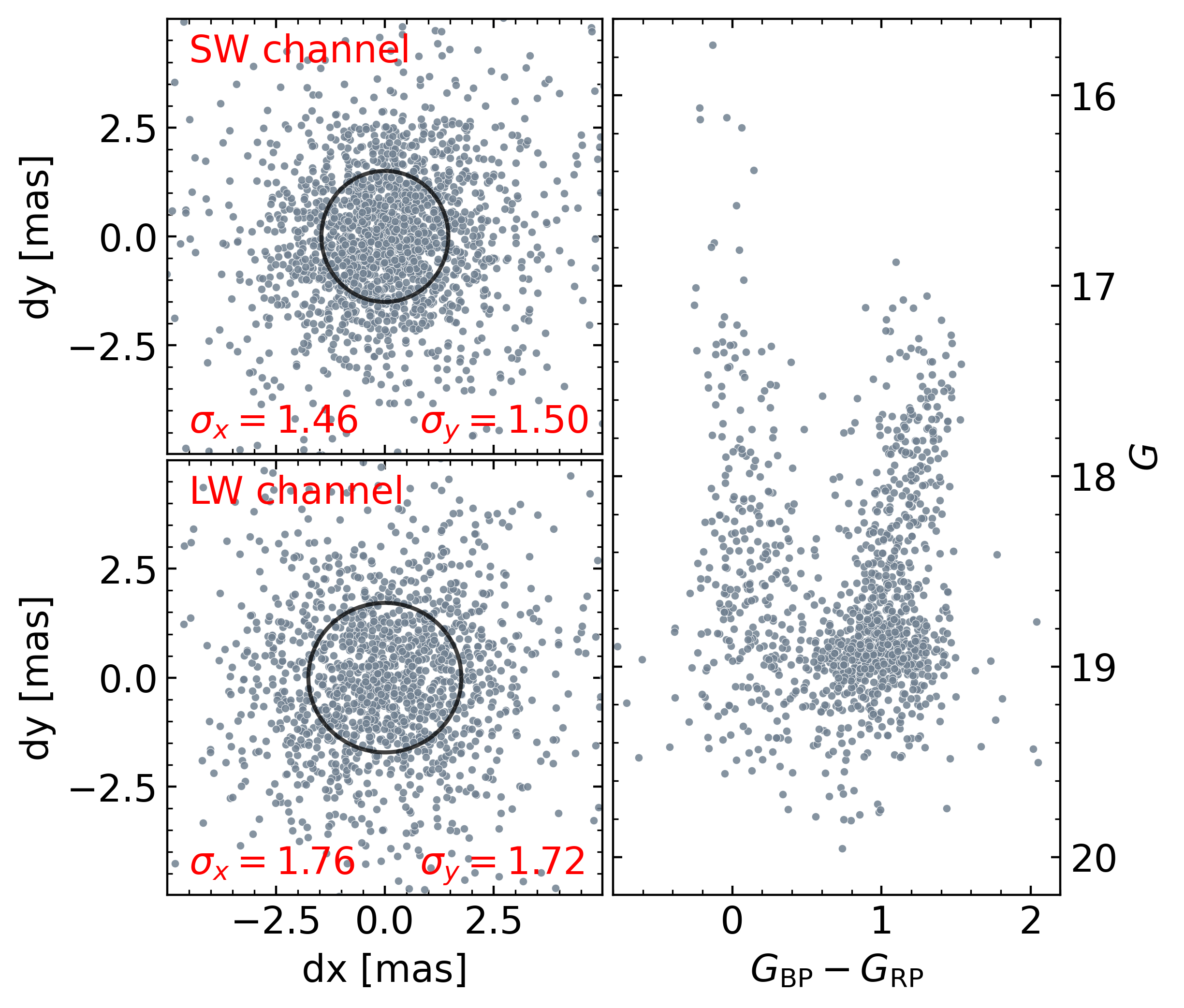}
    \caption{\textit{Left}: positional residuals between the positions measured by us
    and those given by {\it Gaia} DR3 catalogue, for the SW channel (top) and the LW 
    channel (bottom). \textit{Right}: colour-magnitude diagram in the {\it Gaia} filters
    for the common sources.}
    \label{fig:res_dr3}
\end{figure}

Although our formal (internally estimated) errors 
provide uncertainties smaller than 20\,$\mu$as, 
these very likely are underestimates 
of the true errors. 
However, it is not easy to compare 
these corrected positions with other 
catalogues able to reach similar accuracy
for such faint stars. 
The only available is \textit{Gaia}\,DR3 which, 
however, we used to fix the 
linear terms.
Therefore, 
while we will not be able to independently 
test the linear terms of our solution, 
we will nonetheless still be able to test 
the accuracy of the non-linear terms 
of our GD correction. 

Unfortunately, common sources are faint for \textit{Gaia}
and we end up being limited by the errors in the 
\textit{Gaia} catalogue in both positions and motions.
\textit{Gaia}\,DR3 gives positions at the epoch 2016.0, 
while the new \textit{JWST} observations are collected at epoch $\sim$2022.53. 
Given the internal proper-motion dispersion for LMC stars in this field 
(which also has a distribution far from being Gaussian) 
of about 40\,km\,s$^{-1}$ (\citealt{2021jwst..rept....12A}, 
and assuming a distance of 50\,kpc, in 6.53\,years (2022.53-2016.0), we 
expect a dispersion of 1.1\,mas. 

We cross-identified the sources in our SW and LW catalogue  
of a single image (namely \texttt{jw01476001003\_02101\_00001}
in F090W and \texttt{jw01476001001\_02101\_00001}
in F277W) with the {\it Gaia}\,DR3 catalogue,
projected onto the tangent plane of each detector
(as in Sec. \ref{sec:poly1}), and derived the transformations
to bring the positions of our catalogues into the tangent
plane coordinate system.
The residuals between the transformed positions and {\it Gaia}
are shown in Fig.\,\ref{fig:res_dr3} (left panels),
where the dispersions are labelled within each panel as $\sigma$ 
and expressed in mas, for both coordinates. 
The right panel shows the CMD in Gaia bands for sources in common with \textit{JWST}. 
We have tried to extrapolate \textit{Gaia} positions to epoch 2022.53 by employing the 
tabulated \textit{Gaia} proper motions, however, this had the effect to significantly 
enlarging the residual dispersion, making these extrapolations useless.
This is mainly due to both the large errors (0.1-0.5\,mas/yr) on proper motions for faint
sources ($G>17$) in the \textit{Gaia} catalogue, that produce sizeable effects in six years, and the lack of proper motion measurements for very faint stars ($G>20$).

The average observed dispersion for the two coordinates in the SW channel is 1.48\,mas; 
1.74\,mas for LW channel. 
To infer from this the errors of our GD correction, one should subtract in quadrature the 
other contributions that participate to enlarge the dispersion, such as the 
LMC's internal motions ($\sim$1.1\,mas), the errors in the \textit{Gaia} 
catalogue \citep[$\sim$0.25,\,mas][]{2021A&A...649A...1G}, 
and the positioning errors for the bulk of these stars in the \textit{JWST} images 
(0.7\,mas, i.e., the red error bars in Fig.\,\ref{fig:gd_resA1_tab}). 
Summing in quadrature these contributions for the bulk of the 
stars we obtain $\sim$1.3\,mas.
Subtracting in quadrature this value from the observed dispersion for SW, 
we obtain a residual of $\sim$0.6\,mas, which here we entirely ascribe to the 
residual in our GD correction. Doing the same for the LW we obtain again $\sim$0.6\,mas. 

This is a rough estimate for the minimum limit in the accuracy of our GD correction 
and is mainly affected by the strongly non-Gaussian internal proper-motion distributions for 
LMC stars (see Fig.\,15 of \citealt{2021jwst..rept....12A}, and Sect.\,\ref{sec:lmcPMs} of this work). 
Indeed, in next the sections, we will put significantly smaller upper
limits to this estimated accuracy for the here-presented GD correction.

%
\subsection{Internal errors}
\begin{figure}
    \centering
    \includegraphics[width=\columnwidth]{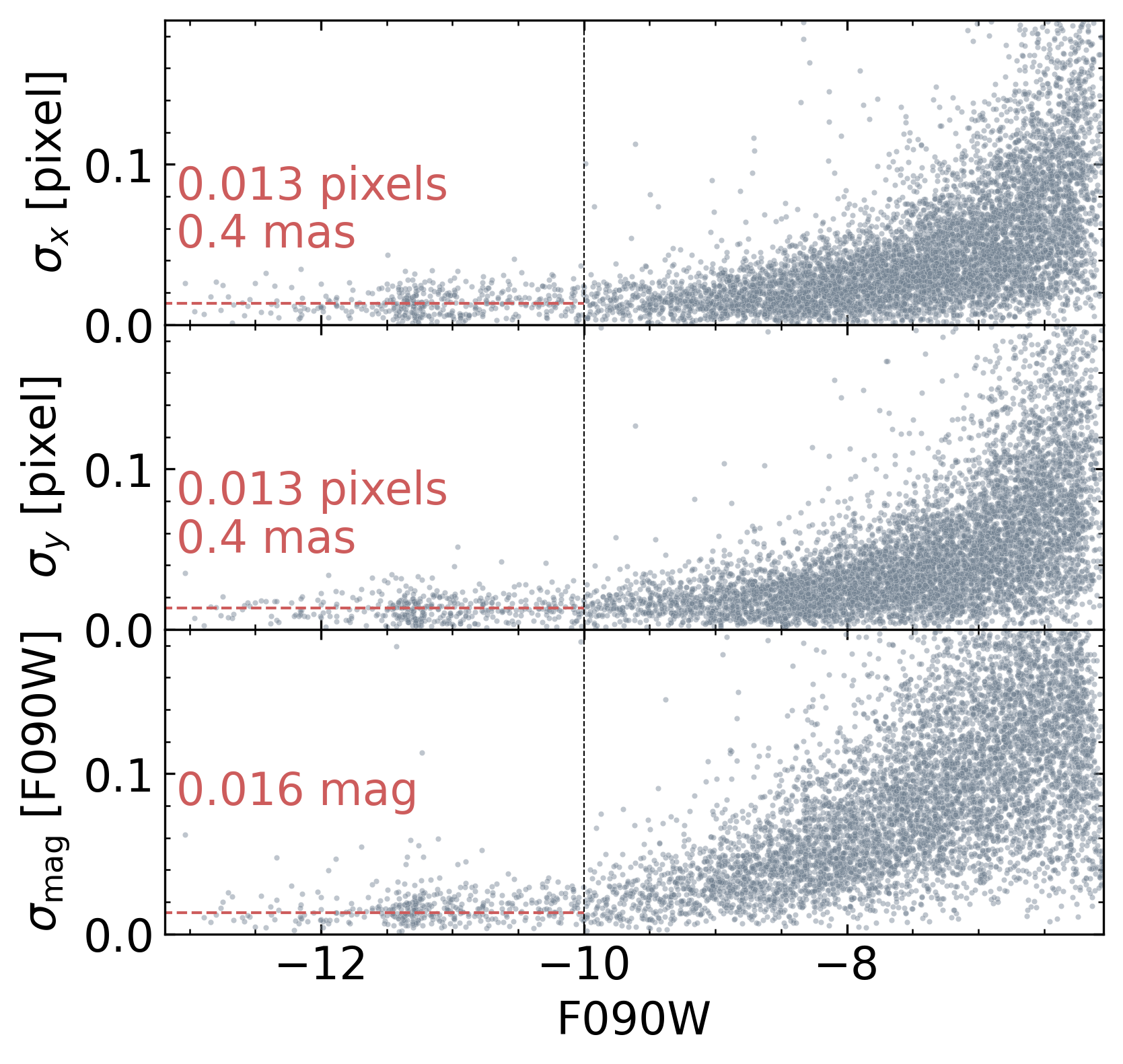}
    \caption{Positional residuals in $x$ and $y$ (top and middle) and magnitude residuals (bottom)
    from the inter-comparison of dithered images in F090W after the GD correction have been applied.
    Red lines indicate the median value of the residuals for well-measured 
    sources ($-13.5<m<-10$, \texttt{QFIT}$<0.2$).}
    \label{fig:sxsy}
\end{figure}
Figures\,\ref{fig:gd_resA1_p1} and \ref{fig:gd_resA1_tab} give positional residuals for the 
bulk of the measured sources in the field (red error bars), here instead, we want to 
show these quantities as a function of the instrumental magnitude for 
the individual sources. 
This is possible by inter-comparing the positions and magnitudes measured employing 
the same filter (9 dithered images), which provide an estimate of the expected r.m.s. 
of the quantities, as measured in a single image, for individual sources. 
In Fig.\,\ref{fig:sxsy}, we show for the case of detector A1 in F090W 
these trends, with median values of 0.013\,pixels (i.e., 0.4\,mas) for the 
1-D positioning, and 16\,milli-mag in the photometry for well-measured sources,
i.e. those with $-13.5<m<-10$ and \texttt{QFIT}$<0.2$.
Similar results are obtained for the other detectors/filters. 
In the following applications and considerations, it is important to distinguish 
the difference between these random errors for individual sources and the systematic 
errors of geometric distortion residuals.

\subsection{Putting detector-based positions into a common reference system}
\label{sec:meta}
In this section we derive the transformations to bring the positions
measured by each detector of a given image into a common reference system
(which arbitrarily we chose to be that of A1).
We remark that 
for investigations 
requiring the most possible accurate differential astrometry,
it is always optimal to compare position measurements as locally as possible, 
provided that there are enough reference sources within the field.
However, some projects might have limited reference objects, and would require 
an understanding of the distortion across the entire NIRCam field of view.
To derive these transformations we exploited 
the {\it Gaia} catalogue. We considered each of the nine dithers separately, 
and we treated every filter independently.
We proceeded as follows:
\begin{itemize}
    \item[--] first, we downloaded a portion of the {\it Gaia} DR3 catalogue
    large enough to cover both modules A and B of the considered exposure;
    \item[--] we projected it onto the plane tangent to the centre of the two modules;
    \item[--] we transformed {\it Gaia} positions into the reference
    system of A1;
    \item[--] finally, we used the {\it Gaia} positions on A1 to derive
    the six parameters transformations to bring all the other detectors on 
    the reference system of A1.
\end{itemize}

At the end of this procedure, for each filter, we have nine transformations for each
detector (one for each of the nine dithers). 
We checked the consistency between the coefficients derived from each
dither, and given the general agreement among them, we computed the final
transformation averaging all the nine estimates.
Furthermore, as the coefficients were compatible even in different filters,
we also averaged the coefficients obtained in the three filters for the SW-channel, 
and averaged those in the two filters for the LW channel, 
resulting in six parameters for the transformation of each detector into a common reference system,
independently of the adopted filter.

In Sect.\,\ref{sec:sum}, we describe the \texttt{Python} software, which we release with this publication, 
that enables users to put all the 10 individual detectors of NIRCam into a common distortion-free meta-chip frame.

\subsection{The absolute scale}

The transformation between A1 and {\it Gaia} let us infer the pixel scale
of our GD-corrected pixel reference system. 
For each of the filters, we observe the nine dithers to agree within few $10^{-5}$, 
with a pixel scale of about $31.23$\,mas/px (see Table \ref{tab:pxsc}).

In the case of {\it HST} the telescope was orbiting at 7\,km\,s$^{-1}$ around the Earth,  
a speed that causes scale variation due to velocity aberration of 
about 7/300\,000 parts, i.e., \textit{also} of few 10$^{-5}$, and every 2\,hours \citep{2003hstc.conf...58C}; 
therefore variations of the same order of what we observe here for \textit{JWST}.
However, unlike {\it HST}, {\it JWST} it is not orbiting at 7 km/s around the Earth. 
Nevertheless, {\it JWST} (as well as \textit{HST}) is still orbiting the Sun with a velocity 
slightly less than 30\,km\,s$^{-1}$, i.e., causing scale variations due to velocity aberration 
of about 30/300\,000, or 1 part in 10\,000, a very sizeable effect, although with a much 
slower $\sim$6\,months time-frame. 
These effects (of 1 part in 10\,000) needs to be properly accounted for in all applications which blindly rely on the 
absolute scale of the telescope (assuming {\it JWST} will prove to have a scale stable down to this level).
For this purpose, the calibration pipeline includes in the header of each image the 
expected velocity aberration scale factor (\texttt{VA\_SCALE}) calculated on the base of the 
expected absolute velocity of the Observatory. 
We note that the values of the \texttt{VA\_SCALE} reported in the header of each image of the here-employed 
1476 data-set, remains well below the few 10$^{-5}$ scatter observed. 
In the lack of other observations, we assumed this to be the limit of the plate-scale stability for \textit{JWST}. 

We are deriving our absolute scales comparing directly to the 
absolute astrometric reference frame of {\it Gaia}\,DR3, therefore, 
to retrieve the true scale of our GD solution we should first divide for the \texttt{VA\_SCALE} factor. 
The results obtained for the average of the scale of detector A1 
compared to {\it Gaia}\,DR3, for all images collected in filters F090W, F150W, and F150W2, 
are shown in Table\,\ref{tab:pxsc}.
The values for each filter are the averaged values obtained from the nine dithers.
Note that the scale for filter F090W is significantly different (at $\sim$14\,$\sigma$) from the one 
for the two filters F150W and F150W2, which are instead marginally consistent ($\sim$2.8\,$\sigma$) 
among them.

Finally, we note that this is the scale to apply to our --here-derived-- GD solution, 
that is normalised to a specific chip location. Other GD solutions might refer to different pixels, and therefore might have slightly different scales. 

\begin{table}
    \centering
    \caption{Mean pixel scale ($S$) and \texttt{VA\_SCALE}.}
    \begin{tabularx}{\columnwidth}{lccc}
        \hline
        \hline
        Filter & $S$[mas/px] & $\sigma_{S}$[mas/px] & $\rm\texttt{VA\_SCALE}-1$ \\
        \hline
        F090W   & 31.23227 & 0.00005 & 3.65488$\times$10$^{-6}$\\
        F150W   & 31.23115 & 0.00006 & 3.64514$\times$10$^{-6}$\\
        F150W2  & 31.23086 & 0.00008 & 3.64028$\times$10$^{-6}$\\
        \hline
    \end{tabularx}
    \label{tab:pxsc}
\end{table}

\section{Colour-magnitude diagrams}
\begin{figure*}
    \centering
    \includegraphics[width=.9\textwidth]{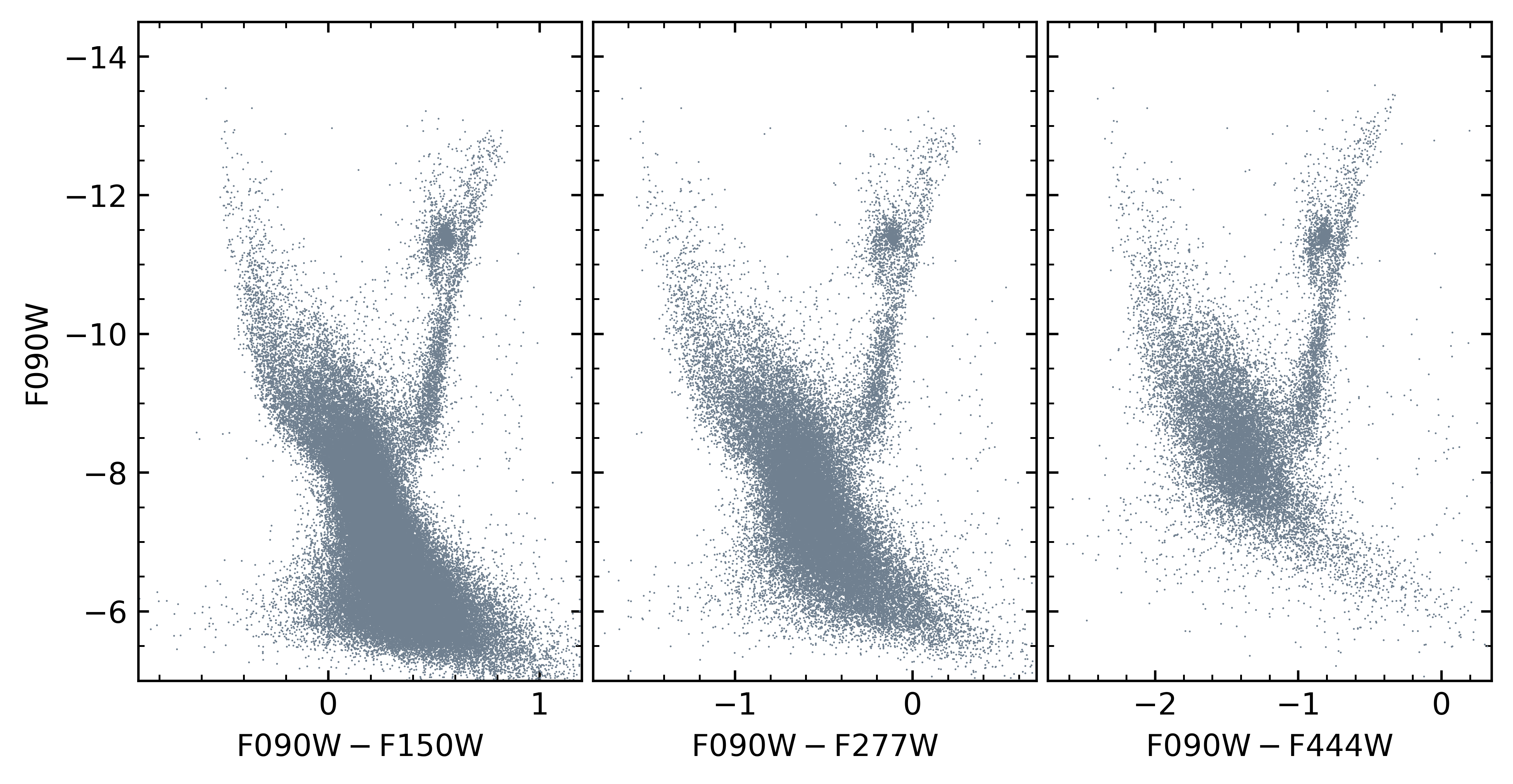}
    \caption{The F090W versus F090W$-$X CMDs, with X\,$=$\,F150W, F277W, F444W, of the stars in the LMC, obtained with the data used in this work.} 
    \label{fig:cmd}
\end{figure*}


The dither pattern of the observations 
provides large overlaps, which in turn,  
allows us to compare the photometry 
obtained from the different detectors of a module 
to register the zero points of the detectors into 
a common photometric reference system.
In the case of the SW channel,
we chose as reference system the first image obtained with detectors A1 and B1, 
and, for each filter and module, 
we transformed the positions and magnitudes of the stars measured in all the images into the 
reference system defined by this first image. 
We do this for each module, separately. 
We averaged the transformed positions and magnitudes of each detector, to 
obtain a more robust catalogue of stars measured in at least three images and we iterated refining 
the transformations by using as reference system the new catalogue containing the mean positions and magnitudes. We report in Table~\ref{tab:zp} the  photometric zero-points of each detector within
each module of the SW channel compared to detectors A1 and B1 (which by definition have null shifts).

Even if the overlap between modules A and B is small, we also were able to measure the zero points 
$\delta$mag=mag[A]-mag[B] 
between the catalogues obtained with the different modules in one filter (it means the zero-point between A1 and B1 in the case of SW channel, and A and B in the case of LW channel). We found $\delta {\rm F090W}=-0.31 \pm 0.06$, $\delta {\rm F150W}=-0.20 \pm 0.07$,  $\delta {\rm F277W}=+0.03 \pm 0.11$, and  $\delta {\rm F444W}=-0.06\pm 0.07$. 

For each filter, we carried out selections by using quality parameters like the photometric RMS and the quality-of-fit, as done in \citetalias{2022MNRAS.517..484N}. Figure~\ref{fig:cmd} shows the F090W versus F090W$-$X instrumental CMDs of the stars in the LMCs observed by NIRCam in the F090W, F150W, F277W, and  F444W filters and that passed the quality selections. The deepest CMD is the F090W$-$F150W one, which reaches two magnitudes below the MS turn-off with a SN$\sim$5; the same signal is reached by the F444W filter two magnitudes brighter, making this filter the shallowest among those used in this work to follow 
the MS stars of the LMC.
\begin{table}
    \centering
    \caption{Relative photometric zero-points for SW and LW channels.}
    \begin{tabularx}{\columnwidth}{XXX}
        \hline
        \hline
        & SW Channel & \\
        \hline
        {\bf Detector } & {\bf F090W} & {\bf F150W} \\
        \hline
        A1 & $+0.00 \pm 0.00$ & $+0.00 \pm 0.00$ \\
        A2 & $+0.06 \pm 0.03$ & $+0.02 \pm 0.02$ \\
        A3 & $+0.06 \pm 0.02$ & $+0.06 \pm 0.02$ \\
        A4 & $+0.00 \pm 0.02$ & $-0.04 \pm 0.02$ \\
        \hline
        B1 & $+0.00 \pm 0.00$ & $+0.00 \pm 0.00$ \\
        B2 & $-0.02 \pm 0.02$ & $-0.01 \pm 0.02$ \\
        B3 & $-0.09 \pm 0.02$ & $-0.06 \pm 0.02$ \\
        B4 & $-0.17 \pm 0.02$ & $-0.18 \pm 0.01$ \\
        \hline
        $\delta$(A1B1) & $-0.31\pm0.06$ & $-0.20\pm0.07$ \\
        \hline
        & LW Channel & \\
        \hline
        {\bf Detector } & {\bf F277W} & {\bf F444W} \\
        \hline
        $\delta$(AB) & $+0.03\pm0.11$ & $-0.06\pm0.07$ \\
        \hline
    \end{tabularx}
    \label{tab:zp}
\end{table}
%
\section{Demonstrative applications}

In this section,  
we demonstrate that applying our just-derived GD correction to positions of sources, 
and comparing these positions with those measured in an earlier archival {\it HST} data set,  
we are able to detect stellar motions at sub-mas level precision. 

To this aim, we considered three applications, sorted by increasing difficulty: 
\textit{(1)} the cluster-field separation in the case of the globular cluster M92;  
\textit{(2)} the estimate of the internal motion of the same cluster; and finally
\textit{(3)} the clear detection of the internal motions in the LMC system, a 
stellar system at $\sim$50\,kpc.

\begin{figure}
    \centering
    \includegraphics[width=.85\columnwidth]{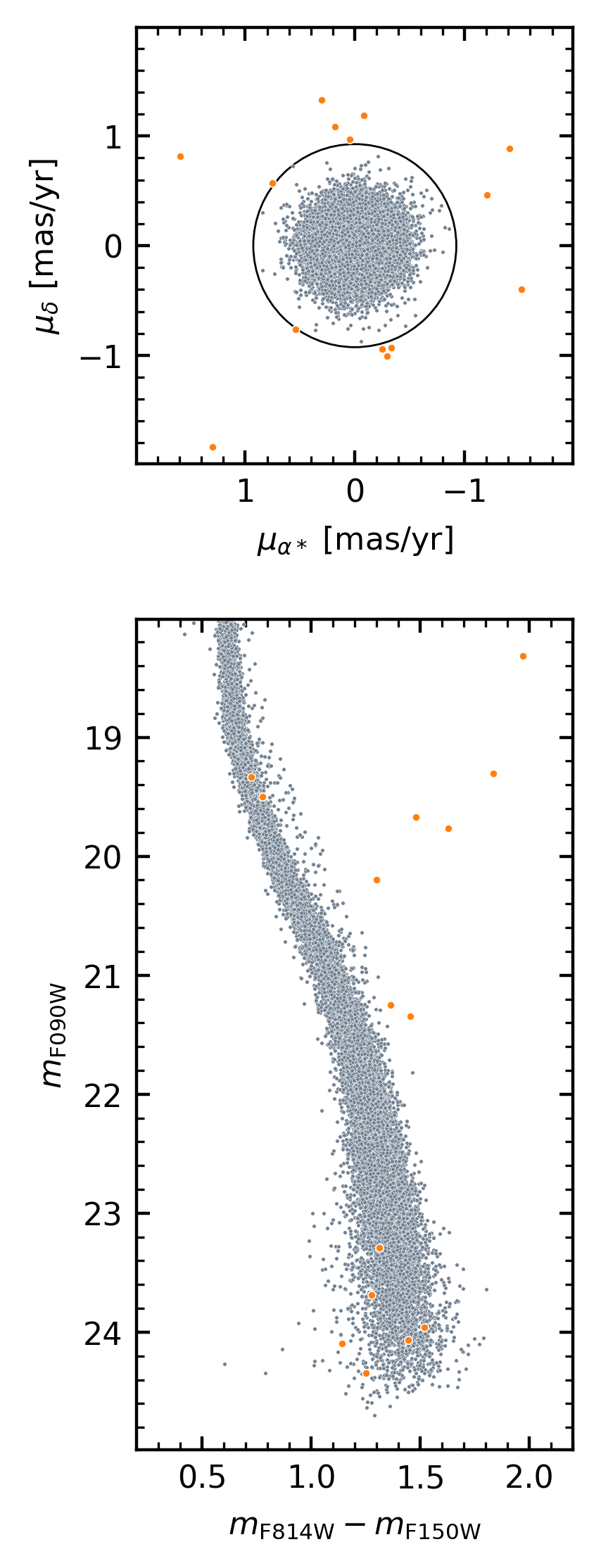} 
    \caption{
    \textit{Top}: vector-point-diagram of proper motions for sources in the common field 
    between images collected with \textit{HST} under program GO-10775, and 
    the available images from \textit{JWST} program ERS-1334.
    A black circle defines our arbitrary criterion to separate members (grey) and field objects (orange).
    \textit{Bottom}: CMD in filters F814W$-$F150W vs.\, F090W; colour code is the same as top panel.  
    } 
    \label{fig:vpd}
\end{figure}

\subsection{Field-object decontamination in M92} 
To compute the displacements of the stars in a field centred in M92, we adopted as 
the first epoch the {\it HST} 
observations collected under programme GO-10775
(PI: Sarajedini, \citealt{2007AJ....133.1658S}, epoch 2006.27),
and as a second epoch the {\it JWST} data from the 
ERS-\href{https://www.stsci.edu/jwst/science-execution/program-information.html?id=1334}{1334}
(PI: Weisz, epoch 2022.47). 
For the first epoch, we used the catalogue obtained by \citet{2018MNRAS.481.3382N}, while for the second epoch we used 
the catalogues obtained in \citetalias{2022MNRAS.517..484N}, corrected by using the GD solution of this work.
We matched the {\it HST} F814W catalogue with the {\it JWST} F090W and F150W catalogues by using 6-parameter global transformations. Sources that moved the least, and by far the large majority, are M92 member stars, 
therefore the zero of the motion coincides with the mean motion of the cluster. 
Top panel of Figure\,\ref{fig:vpd} shows the resulting vector-point diagram (VPD) of the displacements of the stars in $\delta t=16.2$\,yrs. 
We arbitrarily defined as field stars all the sources with a proper motion larger than $\sim 0.9$\,mas/yr (red points), 
which is about 3.5\,$\sigma$ of the internal distribution (see next sect.). 
The bottom panel shows, for the same sources in the VPD, 
the $m_{\rm F090W}$ versus $m_{\rm F814W}-m_{F150W}$ CMD, employing the same symbols and colour codes.

Unfortunately, M92 is not an ideal target for a striking demonstration of the cluster-members field-objects decontamination, mainly because of the extremely sparse density of Galactic and extra-galactic sources in the 
direction of M92, where we count about 15 sources. 
%

\begin{figure}
    \centering
    \includegraphics[width=\columnwidth]{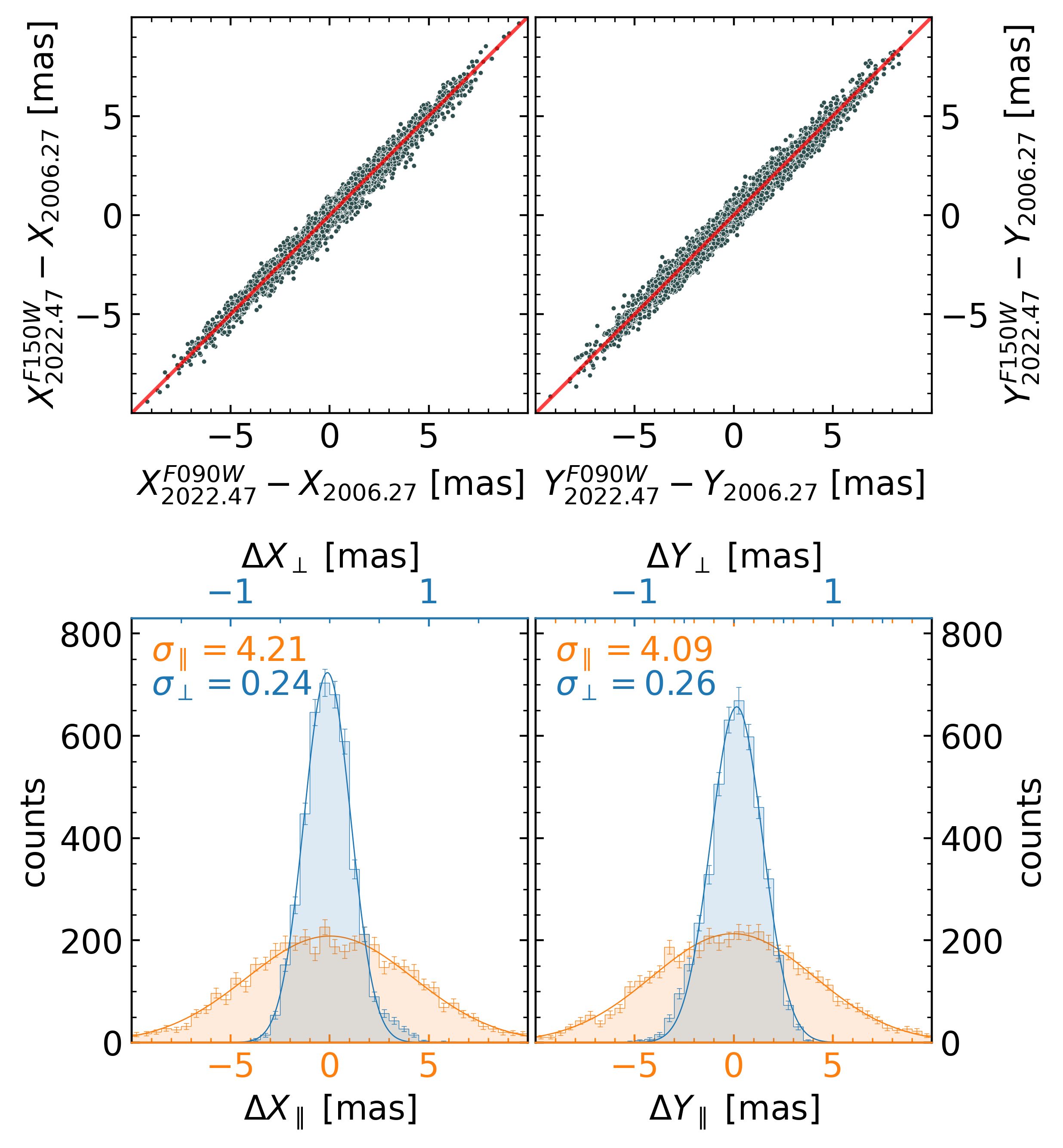}
\caption{
    \textit{(Top:)} Correlation between positional displacements 
    obtained using an \textit{HST} epoch collected in 2006.27 with filter ACS/WFC/F814W, 
    and two different \textit{JWST} data sets, in filters NIRCam/SW/F090W and F150W, both collected in 2022.47.
    The identity is indicated by the red line. 
    $X$-coordinate on the \textit{(left)}, and $Y$ on the \textit{(right)}. 
    \textit{(Bottom:)} Histogram of the displacement distributions along and perpendicular to the identity line.
    Note the different scale for the two quantities, given in the opposite axes (see text). 
    }
    \label{fig:pmm92}
\end{figure}
\subsection{M92 internal dispersion}

The globular cluster M92 (NGC\,6341) is a relatively massive system (3.5$\times$10$^5$\,M$_\odot$)
located at a distance of $\sim$8.5\,kpc and with a half-mass of 4.5\,pc, i.e., 110$^{\prime\prime}$ 
\citep[][hereafter VB21]{2021MNRAS.505.5978V}.
In the radial range explored by the combined \textit{HST}-\textit{JWST} epochs, i.e., 20-100\,arcsec from 
the centre of the cluster, according to the literature, we expect internal-velocity dispersion between 8 and 
5.5\,km\,s$^{-1}$ (i.e., between 0.20 and 0.12\,mas\,yr$^{-1}$, \citetalias{2021MNRAS.505.5978V}). 

In this section, we further test our GD correction 
estimating the internal-proper motion dispersion for M92, 
and in the process we will also obtain a check on the precision of NIRCam astrometry. 

We consider positions $(X,Y)$ measured within \textit{JWST} at epoch 2022.47, in the 
two filters F090W and F150W, separately. 
With those positions, we computed the displacements of sources in F090W and F150W, with respect to 
those measured within \textit{HST} at epoch 2006.27, for each filter separately. 
We selected best-measured sources in all data sets, in the brightest 2\,magnitudes just 
below the saturation and with photometric diagnostics \texttt{QFIT} and r.m.s.\ selected as 
described in Fig\,5 of \citetalias{2022MNRAS.517..484N}. 
We plot the two displacements in top panels of Fig.\,\ref{fig:pmm92}. 
The first epoch is identical for the two computed displacements, 
and the second epoch is essentially \textit{also} the same.
Indeed, F150W images were collected only a few minutes after F090W images, 
which does not change much the time baseline of $\sim$16.2\,years. 
The two displacements correlate with the identity (red line, not a fit) in both 
coordinates ($X$ on the left, $Y$ on the right). 
Assuming Gaussian distributions for both the dispersion along the red line ($\sigma_{\parallel}$), and 
perpendicularly to it ($\sigma_{\rm \perp}$), we can derive crude estimates for both dispersion in the  
M92's internal motions ($\sigma_{\rm intr}$) and for the errors ($\sigma_{\rm err}$). 
Bottom panels in Fig.\,\ref{fig:pmm92} show the histograms for displacements 
along the parallel (in orange) and perpendicular (in blue) to the identity line. 

Any error in the \textit{HST} 2006 epoch has the effect to move a source only along the red line. 
So, the cross dispersion, i.e., perpendicular to the red line, reflects the errors only in the 
two \textit{JWST} epochs.  
Assuming the same dispersion for the two filters, $\sigma_{\rm JWST}$, we can write 
$\sigma_{\rm JWST} = \sigma_{\perp}/\sqrt{2}$.  
Taking the average in $X$ and $Y$ we obtain $\sigma_{\perp}=0.25$\,mas, and therefore 
$\sigma_{\rm JWST} = 0.18$\,mas for the single \textit{JWST} epoch  
(note, dispersion of displacements not of proper motions). 
This is essentially, just another way to put an upper limit to the errors in on our GD correction, 
although, internal to the method.  

As four single \textit{JWST} images participate in the precision of the single filter, 
we can multiply  by a factor $\sqrt{(4-1)}$ to get the precision for the typical star in 
the individual image, about 0.3\,mas, or $\sim$0.01\,pixel; consistent 
with positioning precision for the best measured sources, which means that 
the errors in our GD corrections should be negligible with respect to it.

Now, we try to infer an estimate of the intrinsic proper-motion dispersion of M92 stars
in the region covered by the two epochs. 
Similarly to what was done for the errors, we can assume that 
$\sigma_{\rm obs} = \sigma_{\parallel}/\sqrt{2}$.  
Again, taking the average of $X$ and $Y$ we obtain a $\sigma_{\parallel}=4.15$\,mas, 
and therefore $\sigma_{\rm obs} = 2.93$\,mas. 
To know the intrinsic displacement dispersion we need to subtract in quadrature the 
errors. To the errors this time participate one \textit{JWST} and one \textit{HST} epoch. 
So, we sum in quadrature the errors just derived above $\sigma_{\rm JWST} = 0.18$\,mas, 
and assume \textit{HST} errors from the literature.
For best stars we expect 0.32\,mas \citep[from][]{2011PASP..123..622B}, but as four \textit{HST} images 
from 2006 participate to determine the positions, we take 
$\sigma_{\rm HST} = 0.32$\,mas/$\sqrt{(4-1)}=0.18$\,mas.  
This makes the total errors, sum in quadrature of $\sigma_{\rm JWST}$ and $\sigma_{\rm HST}$,  
amount to $\sigma_{\rm err} = 0.25$\,mas; negligible when compared to $\sigma_{\rm obs}$ 
(as obvious from a glance to Fig.\,\ref{fig:pmm92}).
Nevertheless, the intrinsic dispersion of the observed displacement is 
$\sigma_{\rm intr} = \sqrt{\sigma^2_{\rm obs}-\sigma^2_{\rm err}} = 2.92$\,mas.  

Finally, taking into account the time base-line of 16.2\,yr, 
we derive an estimate for the internal proper motion of M92 stars of 
0.18\,mas\,yr$^{-1}$; consistent with the value found by \citetalias{2021MNRAS.505.5978V} in the core 
(0.2\,mas\,yr$^{-1}$ in the centre, and 0.1\,mas\,yr$^{-1}$ at 100\,arcsec). 
Indeed, our star sample is biased toward the centre, due to the spatial distribution of sources in
a globular cluster. 

The result is even more remarkable, taking into account that in the process 
of deriving displacement, we use a global transformation to transform from \textit{HST} 
into \textit{JWST} master frames, letting us completely at the mercy of residual in 
the geometric distortion of both \textit{JWST} and \textit{ACS} (which are sizable in 
this particular data set, cfr. Sect. 4.3 of \citealt{2021jwst..rept....12A}, 
but thankfully diluted in a 16.2\,yr time baseline). 
This means that by using a local transformations approach
\citep[as described in, e.g.][]{2003AJ....126..247B,2006A&A...454.1029A,2018ApJ...853...86B}
residual errors of various origins, could be suppressed. 
For these reasons, the results presented in this section are even more impressive.

\begin{figure}
    \centering
    \includegraphics[width=\columnwidth]{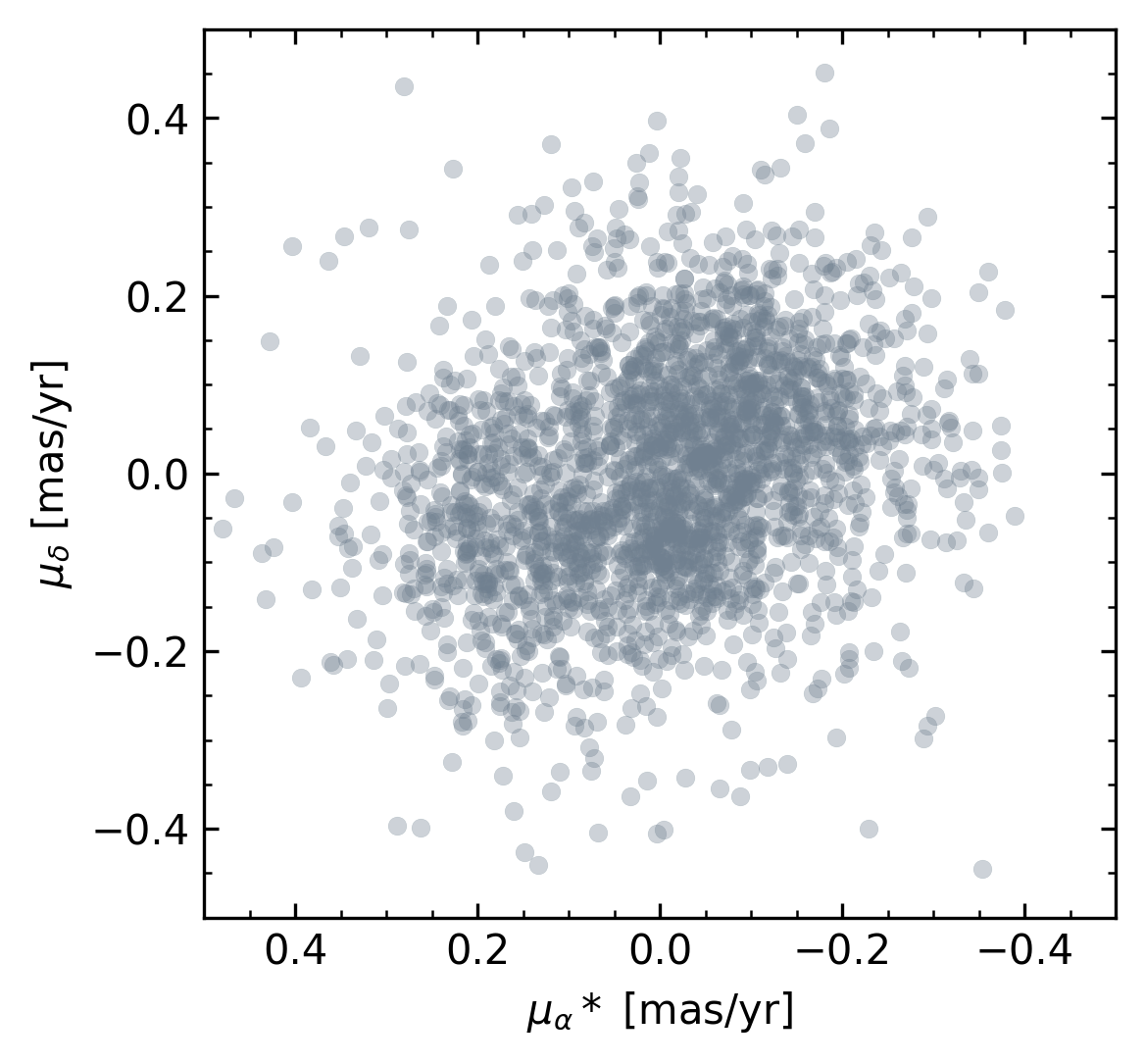} 
    \caption{
Displacements positions for sources in the LMC field as measured in archival {\it HST} images in year 2006.39, 
and the corresponding positions as measured in {\it JWST} in filter F090W at epoch 2022.53.
    } 
    \label{fig:pm}
\end{figure}
\subsection{LMC internal dispersion}
\label{sec:lmcPMs}

To detect the internal proper motion dispersion of LMC stars, we adopted as a 
first epoch the {\it HST} data collected during the calibration program CAL/OTA-10753 (PI: Diaz-Miller).
The data set consists of $5\times 19$~s$+2\times 32$~s$+25\times 343$~s$+10\times 423$~s ACS/WFC 
images in F606W filter and observations were carried out between 25 April and 9 July 2006 (mean epoch $t \sim 2006.39$). 
A catalogue of sources was extracted for each image by using the software \texttt{hst1pass} \citep{2022wfc..rept....5A}. These catalogues were matched with {\it Gaia} DR3 catalogue of the same region by using 6-parameter global transformations to orient and transform all the positions of the stars in the same reference system; the transformed positions were then (3$\sigma$-clipped) averaged to obtain a catalogue of stars with positions referred to the epoch 2006.39.
We performed the same transformations with the {\it JWST} GD corrected catalogues in F090W;
the final product consists of a catalogue with positions corresponding to the epoch 2022.53.
We matched the \textit{HST} ACS/WFC/F606W catalogue from 2006.39 
with the F090W catalogue obtained with {\it JWST} in 2022.53,  
by using 6-parameter global (i.e., not local) transformations. 

The displacements, converted in proper motions assuming a time baseline of $\delta t= 16.14$ yr, 
are shown in Fig.~\ref{fig:pm}. 
Beside the flip of the $\mu_\alpha^{*}$ axis, and the zero of the motions referred to LMC stars rest frame, 
the VPD distribution we obtained employing \textit{JWST} is completely consistent with 
the one characterised in great detail by \cite{2021jwst..rept....12A}, for the same region of the LMC. 
The VPD has the same strongly non-Gaussian distribution in both $\alpha$ and $\delta$, with three-lobed shape, clearly recognisable also in our Fig.~\ref{fig:pm}. This further,  demonstrate that our NIRCam GD correction enables us to obtain high-precision results comparable to what obtainable with \textit{HST}. 

As a final note, a more solid estimate of the internal velocity dispersion 
within LMC would be obtained by performing 
a local transformation approach \citep[as for example in][]{2014MNRAS.439..354B}.

\section{Conclusion}
\label{sec:sum}

In this work, we have exploited {\it JWST} observations of a field
in the LMC and {\it Gaia}\,DR3 to calibrate the geometric distortion of the 
ten NIRCam detectors. 
We exploited the calibrated positions coupling them with
archival \textit{HST} observations to measure the proper 
motions of sources within a field in the core of the Galactic globular cluster M92.
Our measurements were able to clearly disentangle 
field objects from cluster members, and even to measure their internal kinematic.
We also were able to measure the internal dispersion of stars within one extra-galactic system, 
the LMC. 
In all cases, our results are in agreement with the literature and 
in line with state-of-the-art astrometry.  

Indeed, it is worth mentioning that the here-presented GD correction was 
successfully employed  in the recent work by \cite{2023arXiv230204879N}:
where it allowed for the first detection of brown dwarf candidates
in a GC, as result of careful image registration, and accurate proper-motion memberships 
derived by the coupling with existing \textit{HST} archival material collected $\sim$12\,years earlier.

Finally, we publicly release two \texttt{Python} tools to apply our
geometric distortion correction to the raw coordinates
of NIRCam detectors, and to put all the detector-based 
positions into a common, distortion-free, global reference system.
The routine \texttt{raw2cor.py} takes as input a list of raw coordinates,
the module (A or B)
and the detector (1-5), and applies the third-degree polynomial,
the linear terms, and the fine-scale table, giving as output
the corrected coordinates.
The routine \texttt{xy2meta.py} requires the same input
as \texttt{raw2cor.py}, but in addition to the GD corrections
it also applies the transformations to bring all the coordinates
into a common reference system, which are given as output.
These routines are released as electronic material with this paper 
and are also downloadable from the following url: 
\url{https://oapd.inaf.it/bedin/files/PAPERs_eMATERIALs/JWST/Paper_02/Python}.

\section*{Acknowledgements}
The authors MG, DN and LRB acknowledge support by MIUR under PRIN program \#2017Z2HSMF and 
by PRIN-INAF\,2019 under program \#10-Bedin.
This work is based on observations made with the NASA/ESA/CSA James
Webb Space Telescope. The data were obtained from the Mikulski Archive
for Space Telescopes at the Space Telescope Science Institute, which
is operated by the Association of Universities for Research in
Astronomy, Inc., under NASA contract NAS 5-03127 for \textit{JWST}. 
The data employed in this work are associated with 
calibration program CAL-1476 (PI: Boyer) and 
early release science program ERS-1334 (PI: Weisz).  

This research is also based on observations made with the NASA/ESA
{\it Hubble Space Telescope} obtained from the Space Telescope Science
Institute, which is operated by the Association of Universities for
Research in Astronomy, Inc., under NASA contract NAS 5-26555.
These observations are associated with programs CAL/OTA-10753 (PI: Diaz-Miller), GO-10775 (PI:
Sarajedini).

This work has made use of data from the European Space Agency (ESA) mission
{\it Gaia} (\url{https://www.cosmos.esa.int/gaia}), processed by the {\it Gaia}
Data Processing and Analysis Consortium (DPAC,
\url{https://www.cosmos.esa.int/web/gaia/dpac/consortium}).

\bibliography{biblio}

\begin{thebibliography}{}

\bibitem [\protect \citeauthoryear {%
{Anderson}%
}{%
{Anderson}%
}{%
{\protect \APACyear {2022}}%
}]{%
2022wfc..rept....5A}
\APACinsertmetastar {%
2022wfc..rept....5A}%
\begin{APACrefauthors}%
{Anderson}, J.%
\end{APACrefauthors}%
\unskip\
\newblock
\APACrefYearMonthDay{2022}{{\APACmonth{07}}}{},
\newblock
\APACrefbtitle {{One-Pass HST Photometry with hst1pass}.} {{One-Pass HST
  Photometry with hst1pass}.},
\newblock
\APAChowpublished {Instrument Science Report WFC3 2022-5, 55 pages}.
\PrintBackRefs{\CurrentBib}

\bibitem [\protect \citeauthoryear {%
{Anderson}%
, {Bedin}%
, {Piotto}%
, {Yadav}%
\BCBL {}\ \BBA {} {Bellini}%
}{%
{Anderson}%
\ \protect \BOthers {.}}{%
{\protect \APACyear {2006}}%
}]{%
2006A&A...454.1029A}
\APACinsertmetastar {%
2006A&A...454.1029A}%
\begin{APACrefauthors}%
{Anderson}, J.%
, {Bedin}, L\BPBI R.%
, {Piotto}, G.%
, {Yadav}, R\BPBI S.%
\BCBL {}\ \BBA {} {Bellini}, A.%
\end{APACrefauthors}%
\unskip\
\newblock
\APACrefYearMonthDay{2006}{{\APACmonth{08}}}{},
\newblock
\unskip
\newblock
\APACjournalVolNumPages{\aap}{454}{3}{1029-1045}.
\newblock
\begin{APACrefDOI} \doi{10.1051/0004-6361:20065004} \end{APACrefDOI}
\PrintBackRefs{\CurrentBib}

\bibitem [\protect \citeauthoryear {%
{Anderson}%
, {Fall}%
\BCBL {}\ \BBA {} the Astrometry Working~Group%
}{%
{Anderson}%
\ \protect \BOthers {.}}{%
{\protect \APACyear {2021}}%
}]{%
2021jwst..rept....12A}
\APACinsertmetastar {%
2021jwst..rept....12A}%
\begin{APACrefauthors}%
{Anderson}, J.%
, {Fall}, M.%
\BCBL {}\ \BBA {} the Astrometry Working~Group.%
\end{APACrefauthors}%
\unskip\
\newblock
\APACrefYearMonthDay{2021}{{\APACmonth{01}}}{},
\newblock
\APACrefbtitle {{The JWST Calibration Field: Absolute Astrometry and Proper
  Motions with GAIA and a Second HST Epoch}.} {{The JWST Calibration Field:
  Absolute Astrometry and Proper Motions with GAIA and a Second HST Epoch}.},
\newblock
\APAChowpublished {JWST Technical Report JWST-STScI-007716, SM-12, 36 pages}.
\PrintBackRefs{\CurrentBib}

\bibitem [\protect \citeauthoryear {%
{Anderson}%
\ \BBA {} {King}%
}{%
{Anderson}%
\ \BBA {} {King}%
}{%
{\protect \APACyear {2003}}%
}]{%
2003PASP..115..113A}
\APACinsertmetastar {%
2003PASP..115..113A}%
\begin{APACrefauthors}%
{Anderson}, J.%
\BCBT {}\ \BBA {} {King}, I\BPBI R.%
\end{APACrefauthors}%
\unskip\
\newblock
\APACrefYearMonthDay{2003}{{\APACmonth{01}}}{},
\newblock
\unskip
\newblock
\APACjournalVolNumPages{\pasp}{115}{803}{113-131}.
\newblock
\begin{APACrefDOI} \doi{10.1086/345491} \end{APACrefDOI}
\PrintBackRefs{\CurrentBib}

\bibitem [\protect \citeauthoryear {%
{Bedin}%
, {Piotto}%
, {King}%
\BCBL {}\ \BBA {} {Anderson}%
}{%
{Bedin}%
\ \protect \BOthers {.}}{%
{\protect \APACyear {2003}}%
}]{%
2003AJ....126..247B}
\APACinsertmetastar {%
2003AJ....126..247B}%
\begin{APACrefauthors}%
{Bedin}, L\BPBI R.%
, {Piotto}, G.%
, {King}, I\BPBI R.%
\BCBL {}\ \BBA {} {Anderson}, J.%
\end{APACrefauthors}%
\unskip\
\newblock
\APACrefYearMonthDay{2003}{{\APACmonth{07}}}{},
\newblock
\unskip
\newblock
\APACjournalVolNumPages{\aj}{126}{1}{247-254}.
\newblock
\begin{APACrefDOI} \doi{10.1086/375646} \end{APACrefDOI}
\PrintBackRefs{\CurrentBib}

\bibitem [\protect \citeauthoryear {%
{Bedin}%
\ \protect \BOthers {.}}{%
{Bedin}%
\ \protect \BOthers {.}}{%
{\protect \APACyear {2014}}%
}]{%
2014MNRAS.439..354B}
\APACinsertmetastar {%
2014MNRAS.439..354B}%
\begin{APACrefauthors}%
{Bedin}, L\BPBI R.%
, {Ruiz-Lapuente}, P.%
, {Gonz{\'a}lez Hern{\'a}ndez}, J\BPBI I.%
, {Canal}, R.%
, {Filippenko}, A\BPBI V.%
\BCBL {}\ \BBA {} {Mendez}, J.%
\end{APACrefauthors}%
\unskip\
\newblock
\APACrefYearMonthDay{2014}{{\APACmonth{03}}}{},
\newblock
\unskip
\newblock
\APACjournalVolNumPages{\mnras}{439}{1}{354-371}.
\newblock
\begin{APACrefDOI} \doi{10.1093/mnras/stt2460} \end{APACrefDOI}
\PrintBackRefs{\CurrentBib}

\bibitem [\protect \citeauthoryear {%
{Bellini}%
, {Anderson}%
\BCBL {}\ \BBA {} {Bedin}%
}{%
{Bellini}%
\ \protect \BOthers {.}}{%
{\protect \APACyear {2011}}%
}]{%
2011PASP..123..622B}
\APACinsertmetastar {%
2011PASP..123..622B}%
\begin{APACrefauthors}%
{Bellini}, A.%
, {Anderson}, J.%
\BCBL {}\ \BBA {} {Bedin}, L\BPBI R.%
\end{APACrefauthors}%
\unskip\
\newblock
\APACrefYearMonthDay{2011}{{\APACmonth{05}}}{},
\newblock
\unskip
\newblock
\APACjournalVolNumPages{\pasp}{123}{903}{622}.
\newblock
\begin{APACrefDOI} \doi{10.1086/659878} \end{APACrefDOI}
\PrintBackRefs{\CurrentBib}

\bibitem [\protect \citeauthoryear {%
{Bellini}%
\ \BBA {} {Bedin}%
}{%
{Bellini}%
\ \BBA {} {Bedin}%
}{%
{\protect \APACyear {2009}}%
}]{%
2009PASP..121.1419B}
\APACinsertmetastar {%
2009PASP..121.1419B}%
\begin{APACrefauthors}%
{Bellini}, A.%
\BCBT {}\ \BBA {} {Bedin}, L\BPBI R.%
\end{APACrefauthors}%
\unskip\
\newblock
\APACrefYearMonthDay{2009}{{\APACmonth{12}}}{},
\newblock
\unskip
\newblock
\APACjournalVolNumPages{\pasp}{121}{886}{1419}.
\newblock
\begin{APACrefDOI} \doi{10.1086/649061} \end{APACrefDOI}
\PrintBackRefs{\CurrentBib}

\bibitem [\protect \citeauthoryear {%
{Bellini}%
\ \protect \BOthers {.}}{%
{Bellini}%
\ \protect \BOthers {.}}{%
{\protect \APACyear {2018}}%
}]{%
2018ApJ...853...86B}
\APACinsertmetastar {%
2018ApJ...853...86B}%
\begin{APACrefauthors}%
{Bellini}, A.%
, {Libralato}, M.%
, {Bedin}, L\BPBI R.%
\ et al.\end{APACrefauthors}%
\unskip\
\newblock
\APACrefYearMonthDay{2018}{{\APACmonth{01}}}{},
\newblock
\unskip
\newblock
\APACjournalVolNumPages{\apj}{853}{1}{86}.
\newblock
\begin{APACrefDOI} \doi{10.3847/1538-4357/aaa3ec} \end{APACrefDOI}
\PrintBackRefs{\CurrentBib}

\bibitem [\protect \citeauthoryear {%
{Boyer}%
\ \protect \BOthers {.}}{%
{Boyer}%
\ \protect \BOthers {.}}{%
{\protect \APACyear {2011}}%
}]{%
2011AJ....142..103B}
\APACinsertmetastar {%
2011AJ....142..103B}%
\begin{APACrefauthors}%
{Boyer}, M\BPBI L.%
, {Srinivasan}, S.%
, {van Loon}, J\BPBI T.%
\ et al.\end{APACrefauthors}%
\unskip\
\newblock
\APACrefYearMonthDay{2011}{{\APACmonth{10}}}{},
\newblock
\unskip
\newblock
\APACjournalVolNumPages{\aj}{142}{4}{103}.
\newblock
\begin{APACrefDOI} \doi{10.1088/0004-6256/142/4/103} \end{APACrefDOI}
\PrintBackRefs{\CurrentBib}

\bibitem [\protect \citeauthoryear {%
{Cox}%
\ \BBA {} {Gilliland}%
}{%
{Cox}%
\ \BBA {} {Gilliland}%
}{%
{\protect \APACyear {2003}}%
}]{%
2003hstc.conf...58C}
\APACinsertmetastar {%
2003hstc.conf...58C}%
\begin{APACrefauthors}%
{Cox}, C.%
\BCBT {}\ \BBA {} {Gilliland}, R\BPBI L.%
\end{APACrefauthors}%
\unskip\
\newblock
\APACrefYearMonthDay{2003}{{\APACmonth{01}}}{},
\newblock
{\BBOQ}\APACrefatitle {{The Effect of Velocity Aberration on ACS Image
  Processing}} {{The Effect of Velocity Aberration on ACS Image
  Processing}}.{\BBCQ}
\newblock
\BIn{} \APACrefbtitle {HST Calibration Workshop : Hubble after the Installation
  of the ACS and the NICMOS Cooling System} {HST Calibration Workshop : Hubble
  after the Installation of the ACS and the NICMOS Cooling System}\ \BPG~58.
\PrintBackRefs{\CurrentBib}

\bibitem [\protect \citeauthoryear {%
{Gaia Collaboration}%
\ \protect \BOthers {.}}{%
{Gaia Collaboration}%
\ \protect \BOthers {.}}{%
{\protect \APACyear {2021}}%
}]{%
2021A&A...649A...1G}
\APACinsertmetastar {%
2021A&A...649A...1G}%
\begin{APACrefauthors}%
{Gaia Collaboration}%
, {Brown}, A\BPBI G\BPBI A.%
, {Vallenari}, A.%
\ et al.\end{APACrefauthors}%
\unskip\
\newblock
\APACrefYearMonthDay{2021}{{\APACmonth{05}}}{},
\newblock
\unskip
\newblock
\APACjournalVolNumPages{\aap}{649}{}{A1}.
\newblock
\begin{APACrefDOI} \doi{10.1051/0004-6361/202039657} \end{APACrefDOI}
\PrintBackRefs{\CurrentBib}

\bibitem [\protect \citeauthoryear {%
{Gaia Collaboration}%
\ \protect \BOthers {.}}{%
{Gaia Collaboration}%
\ \protect \BOthers {.}}{%
{\protect \APACyear {2022}}%
}]{%
2022arXiv220800211G}
\APACinsertmetastar {%
2022arXiv220800211G}%
\begin{APACrefauthors}%
{Gaia Collaboration}%
, {Vallenari}, A.%
, {Brown}, A\BPBI G\BPBI A.%
\ et al.\end{APACrefauthors}%
\unskip\
\newblock
\APACrefYearMonthDay{2022}{{\APACmonth{07}}}{},
\newblock
\unskip
\newblock
\APACjournalVolNumPages{arXiv e-prints}{}{}{arXiv:2208.00211}.
\PrintBackRefs{\CurrentBib}

\bibitem [\protect \citeauthoryear {%
{Kerber}%
\ \protect \BOthers {.}}{%
{Kerber}%
\ \protect \BOthers {.}}{%
{\protect \APACyear {2019}}%
}]{%
2019MNRAS.484.5530K}
\APACinsertmetastar {%
2019MNRAS.484.5530K}%
\begin{APACrefauthors}%
{Kerber}, L\BPBI O.%
, {Libralato}, M.%
, {Souza}, S\BPBI O.%
\ et al.\end{APACrefauthors}%
\unskip\
\newblock
\APACrefYearMonthDay{2019}{{\APACmonth{04}}}{},
\newblock
\unskip
\newblock
\APACjournalVolNumPages{\mnras}{484}{4}{5530-5550}.
\newblock
\begin{APACrefDOI} \doi{10.1093/mnras/stz003} \end{APACrefDOI}
\PrintBackRefs{\CurrentBib}

\bibitem [\protect \citeauthoryear {%
{Libralato}%
\ \protect \BOthers {.}}{%
{Libralato}%
\ \protect \BOthers {.}}{%
{\protect \APACyear {2015}}%
}]{%
2015MNRAS.450.1664L}
\APACinsertmetastar {%
2015MNRAS.450.1664L}%
\begin{APACrefauthors}%
{Libralato}, M.%
, {Bellini}, A.%
, {Bedin}, L\BPBI R.%
\ et al.\end{APACrefauthors}%
\unskip\
\newblock
\APACrefYearMonthDay{2015}{{\APACmonth{06}}}{},
\newblock
\unskip
\newblock
\APACjournalVolNumPages{\mnras}{450}{2}{1664-1673}.
\newblock
\begin{APACrefDOI} \doi{10.1093/mnras/stv674} \end{APACrefDOI}
\PrintBackRefs{\CurrentBib}

\bibitem [\protect \citeauthoryear {%
{Libralato}%
\ \protect \BOthers {.}}{%
{Libralato}%
\ \protect \BOthers {.}}{%
{\protect \APACyear {2014}}%
}]{%
2014A&A...563A..80L}
\APACinsertmetastar {%
2014A&A...563A..80L}%
\begin{APACrefauthors}%
{Libralato}, M.%
, {Bellini}, A.%
, {Bedin}, L\BPBI R.%
, {Piotto}, G.%
, {Platais}, I.%
, {Kissler-Patig}, M.%
\BCBL {}\ \BBA {} {Milone}, A\BPBI P.%
\end{APACrefauthors}%
\unskip\
\newblock
\APACrefYearMonthDay{2014}{{\APACmonth{03}}}{},
\newblock
\unskip
\newblock
\APACjournalVolNumPages{\aap}{563}{}{A80}.
\newblock
\begin{APACrefDOI} \doi{10.1051/0004-6361/201322059} \end{APACrefDOI}
\PrintBackRefs{\CurrentBib}

\bibitem [\protect \citeauthoryear {%
{Nardiello}%
\ \protect \BOthers {.}}{%
{Nardiello}%
\ \protect \BOthers {.}}{%
{\protect \APACyear {2022}}%
}]{%
2022MNRAS.517..484N}
\APACinsertmetastar {%
2022MNRAS.517..484N}%
\begin{APACrefauthors}%
{Nardiello}, D.%
, {Bedin}, L\BPBI R.%
, {Burgasser}, A.%
, {Salaris}, M.%
, {Cassisi}, S.%
, {Griggio}, M.%
\BCBL {}\ \BBA {} {Scalco}, M.%
\end{APACrefauthors}%
\unskip\
\newblock
\APACrefYearMonthDay{2022}{{\APACmonth{11}}}{},
\newblock
\unskip
\newblock
\APACjournalVolNumPages{\mnras}{517}{1}{484-497}.
\newblock
\begin{APACrefDOI} \doi{10.1093/mnras/stac2659} \end{APACrefDOI}
\PrintBackRefs{\CurrentBib}

\bibitem [\protect \citeauthoryear {%
{Nardiello}%
, {Griggio}%
\BCBL {}\ \BBA {} {Bedin}%
}{%
{Nardiello}%
\ \protect \BOthers {.}}{%
{\protect \APACyear {2023}}%
}]{%
2023arXiv230204879N}
\APACinsertmetastar {%
2023arXiv230204879N}%
\begin{APACrefauthors}%
{Nardiello}, D.%
, {Griggio}, M.%
\BCBL {}\ \BBA {} {Bedin}, L\BPBI R.%
\end{APACrefauthors}%
\unskip\
\newblock
\APACrefYearMonthDay{2023}{{\APACmonth{02}}}{},
\newblock
\unskip
\newblock
\APACjournalVolNumPages{arXiv e-prints}{}{}{arXiv:2302.04879}.
\newblock
\begin{APACrefDOI} \doi{10.48550/arXiv.2302.04879} \end{APACrefDOI}
\PrintBackRefs{\CurrentBib}

\bibitem [\protect \citeauthoryear {%
{Nardiello}%
\ \protect \BOthers {.}}{%
{Nardiello}%
\ \protect \BOthers {.}}{%
{\protect \APACyear {2018}}%
}]{%
2018MNRAS.481.3382N}
\APACinsertmetastar {%
2018MNRAS.481.3382N}%
\begin{APACrefauthors}%
{Nardiello}, D.%
, {Libralato}, M.%
, {Piotto}, G.%
\ et al.\end{APACrefauthors}%
\unskip\
\newblock
\APACrefYearMonthDay{2018}{{\APACmonth{12}}}{},
\newblock
\unskip
\newblock
\APACjournalVolNumPages{\mnras}{481}{3}{3382-3393}.
\newblock
\begin{APACrefDOI} \doi{10.1093/mnras/sty2515} \end{APACrefDOI}
\PrintBackRefs{\CurrentBib}

\bibitem [\protect \citeauthoryear {%
{Sarajedini}%
\ \protect \BOthers {.}}{%
{Sarajedini}%
\ \protect \BOthers {.}}{%
{\protect \APACyear {2007}}%
}]{%
2007AJ....133.1658S}
\APACinsertmetastar {%
2007AJ....133.1658S}%
\begin{APACrefauthors}%
{Sarajedini}, A.%
, {Bedin}, L\BPBI R.%
, {Chaboyer}, B.%
\ et al.\end{APACrefauthors}%
\unskip\
\newblock
\APACrefYearMonthDay{2007}{{\APACmonth{04}}}{},
\newblock
\unskip
\newblock
\APACjournalVolNumPages{\aj}{133}{4}{1658-1672}.
\newblock
\begin{APACrefDOI} \doi{10.1086/511979} \end{APACrefDOI}
\PrintBackRefs{\CurrentBib}

\bibitem [\protect \citeauthoryear {%
{Vasiliev}%
\ \BBA {} {Baumgardt}%
}{%
{Vasiliev}%
\ \BBA {} {Baumgardt}%
}{%
{\protect \APACyear {2021}}%
}]{%
2021MNRAS.505.5978V}
\APACinsertmetastar {%
2021MNRAS.505.5978V}%
\begin{APACrefauthors}%
{Vasiliev}, E.%
\BCBT {}\ \BBA {} {Baumgardt}, H.%
\end{APACrefauthors}%
\unskip\
\newblock
\APACrefYearMonthDay{2021}{{\APACmonth{08}}}{},
\newblock
\unskip
\newblock
\APACjournalVolNumPages{\mnras}{505}{4}{5978-6002}.
\newblock
\begin{APACrefDOI} \doi{10.1093/mnras/stab1475} \end{APACrefDOI}
\PrintBackRefs{\CurrentBib}

\end{thebibliography}

\end{document}